\documentclass{aa} 
\usepackage{txfonts}
\usepackage{natbib}
\usepackage{graphicx}
\usepackage{subfig}
\usepackage{color}
\usepackage{amssymb}
\bibpunct{(}{)}{;}{a}{}{,}

\begin{document}

\title{CH abundance gradient in TMC-1\thanks{This publication is based on observations with the 100-m telescope of the Max-Planck-Institute f\"{u}r Radioastronomie (MPIfR) at Effelsberg.}}
\author{A. Suutarinen\inst{1}
\and W.D. Geppert\inst{2}
\and J. Harju\inst{1,3}
\and A. Heikkil\"{a}\inst{4}
\and S. Hotzel\inst{3,5}
\and M. Juvela\inst{1,3}
\and T. J. Millar\inst{6}
\and C. Walsh\inst{6}
\and J.G.A Wouterloot\inst{7}}

\offprints{A. Suutarinen}

\institute{Department of Physics, P.O.Box 64, FI-00014 University of Helsinki,Finland
\and Department of Physics, AlbaNova, S-10691 Stockholm, Sweden
\and Observatory, University of Helsinki, Finland
\and Onsala Space Observatory, S-43992 Onsala, Sweden
\and GRS mbH, D-50667 Cologne, Germany
\and Astrophysics Research Centre, School of Mathematics and Physics, Queen's University Belfast, University Road, Belfast, BT7 1NN, UK
\and Joint Astronomy Centre, Hilo, HI-96720, USA
}

\date{Received 05/11/2010 / Accepted 14/05/2011}
\abstract
{
}
{
The aim of this study is to examine if the well-known chemical
gradient in \object{TMC-1} is reflected in the amount of rudimentary forms of carbon
available in the gas-phase. As a tracer we use the CH radical 
which is supposed to be well correlated with carbon atoms and simple
hydrocarbon ions. 
}
{
We observed the 9-cm $\Lambda$-doubling lines of CH along
the dense filament of \object{TMC-1}. The CH column densities were compared
with the total H$_2$ column densities derived using the 2MASS NIR data
and previously published SCUBA maps and with OH column densities derived
using previous observations with Effelsberg. We also modelled 
the chemical evolution of \object{TMC-1} adopting physical conditions 
typical of dark clouds
using the UMIST Database for Astrochemistry gas-phase reaction 
network to aid the interpretation of the observed OH/CH abundance ratios.
} 
{
  The CH column density has a clear peak in the vicinity of the
  cyanopolyyne maximum of \object{TMC-1}. The fractional CH abundance relative
  to H$_2$ increases steadily from the northwestern end of the
  filament where it lies around $1.0\,10^{-8}$, to the southeast where
  it reaches a value of $~2.0\,10^{-8}$. The OH and CH column
  densities are well correlated, and we obtained OH/CH abundance
  ratios of $\sim 16-20$.  These values are clearly larger than what
  has been measured recently in diffuse interstellar gas and is likely
  to be related to C to CO conversion at higher densities.  The good
  correlation between CH and OH can be explained by similar production
  and destruction pathways.  We suggest that the observed CH and OH
  abundance gradients are mainly due to enhanced abundances in a
  low-density envelope which becomes more prominent in the
  southeastern part and seems to continue beyond the dense filament.
}
{
An extensive envelope probably signifies an early
stage of dynamical evolution, and conforms with the detection of a
large CH abundance in the southeastern part of the cloud. The implied 
presence of other simple forms of carbon in the 
gas phase provides a natural explanation for the 
observation of ``early-type'' molecules in this region.
} 
\maketitle

\section{Introduction}

The 9-cm $\Lambda$-doubling lines of the
methylidyne radical CH have been commonly used to
trace low density gas in diffuse clouds and at the edges of dense
clouds. Observational surveys have demonstrated a linear correlation
between the CH column density and visual extiction (and thereby also
$N({\rm H_2})$) in low-extinction molecular clouds
\citep[e.g.][]{Hjalmarson77,Mattila86,Magnani93,Magnani05}.  On the
other hand, CH is likely to be critically dependent on the presence of
free carbon in the gas phase.  The modelling results of
\cite{Flower94} predict that CH (or CN) can trace atomic carbon
content even in the cloud interiors.  While the fine structure lines
of \ion{C}{I} become easily optically thick and dominated by the outer layers
of the cloud, the ground state CH lines are optically thin. Therefore one can hope
to ``see'' through the cloud in these lines, even if the cloud edges
may have a significant contribution, and the CH abundance is likely to
vary as a function of density.

We have made observations of the $\Lambda$-doubling transitions of 
CH in its ground state along the dense filament of
Taurus Molecular Cloud-1 (\object{TMC-1}).  With the aid of these observations
we examine the predicted correlation of atomic carbon and
cyanopolyynes and other unsaturated carbon-chain molecules in \object{TMC-1}.

According to most chemistry models adopted for \object{TMC-1} the
cyanopolyyne (CP) peak in its southeastern end should represent an
early stage of chemical evolution and have a high concentration of carbon
in the form of C or C$^+$ \citep[e.g.][]{Hirahara92,Hartquist96,Pratap97}.
Alternative views have been presented by \cite{Hartquist96} who
discuss the possibility that the CP peak is highly depleted and
represents a chemically more advanced stage than the NH$_3$ peak, and
by \cite{Markwick00,Markwick01} who suggest that carbon chemistry at
the CP peak has been revived by evaporation of carbon bearing mantle
species (CH$_4$, C$_2$H$_2$, C$_2$H$_4$) following grain-grain collisions 
induced by the ion-neutral slip produced by the passge of an Alfv{\'e}n wave.
  In this model, complex
organic species are not necessarily coincident with the most
rudimentary forms of carbon.

Previous \ion{C}{I} 492 GHz fine structure line observations of
\cite{Schilke95} and \cite{Maezawa99} indicate high
abundances neutral carbon in the direction of TMC-1. In their mapping
of Heiles Cloud 2 (including TMC-1), Maezawa et al. estimate that the
\ion{C}{I}/CO abundance ratio towards the CP peak is about 0.1, and that the
ratio increases to $\sim1$ in the "\ion{C}{I} peak" on the southern side of
the cloud. These estimates are, however, uncertain because of the
unknown optical thickness and excitation temperature of the 492 GHz \ion{C}{I}
line. The TMC-1 ridge is not visible on the \ion{C}{I} map which
is dominated by emission from more diffuse gas.

In the present paper, we derive the fractional abundances of CH along
an axis aligned with the dense filament of \object{TMC-1} and going through the
CP peak. The molecular hydrogen column densities, $N({\rm H_2})$,
required for this purpose, are estimated by combining an extinction
map derived from the 2MASS data with JCMT/SCUBA dust continuum maps at
850 and 450 $\mu$m presented in \cite{Nutter08}.  The CH observations
are described in Sect. 2, and in Sect. 3 the CH and H$_2$ column
densities are derived. In Sect. 4 we compare the new data with
previous observations, and discuss the chemical implications of the
obtained CH abundances. Finally, in Sect. 5 we summarize our
conclusions.

\section{Observations}
\begin{figure*}
  \subfloat[][]{\label{fig:SCUBAmap}\includegraphics[width=8.5cm]{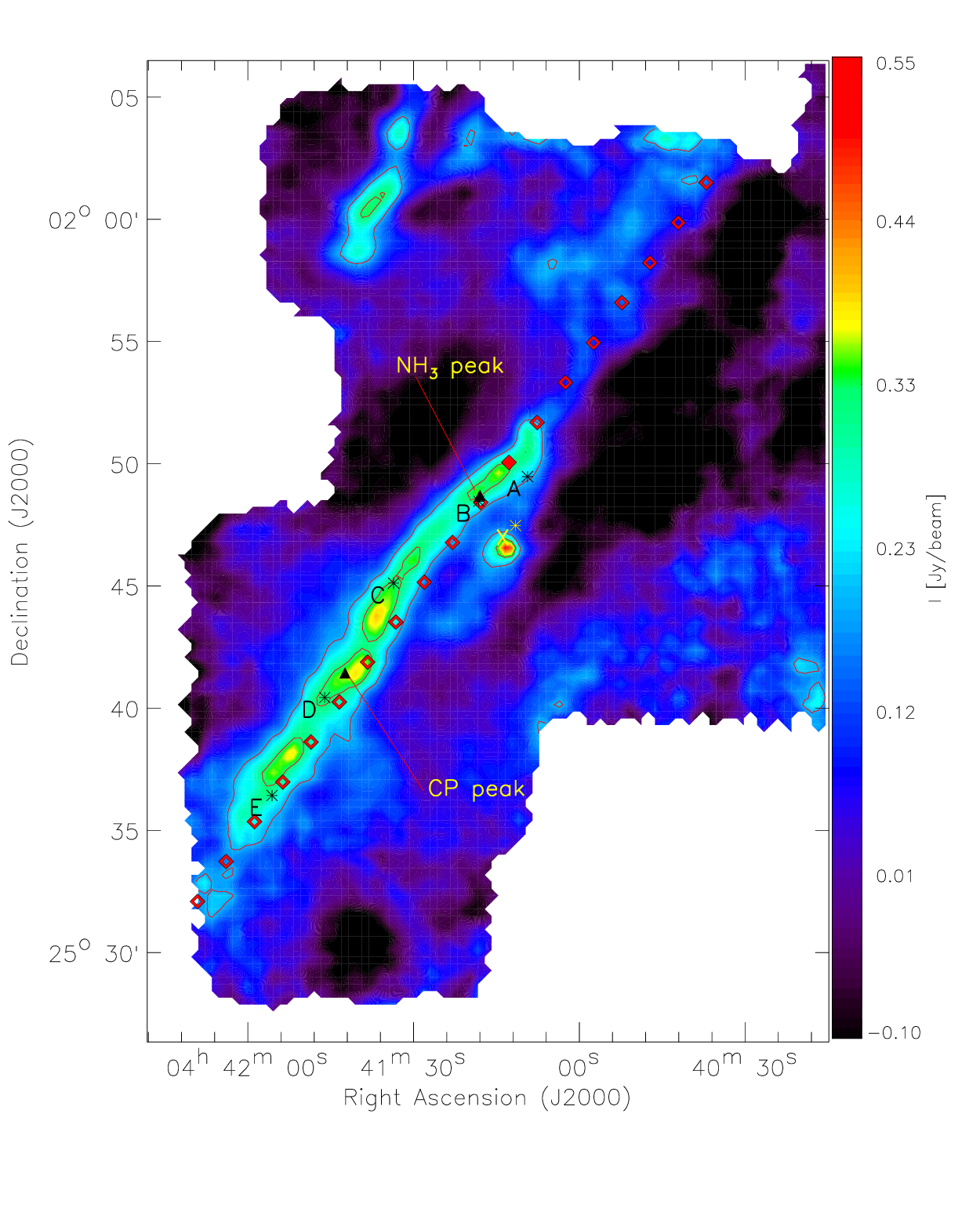}}
  \subfloat[][]{\label{fig:SCUBAmap228}\includegraphics[width=8.5cm]{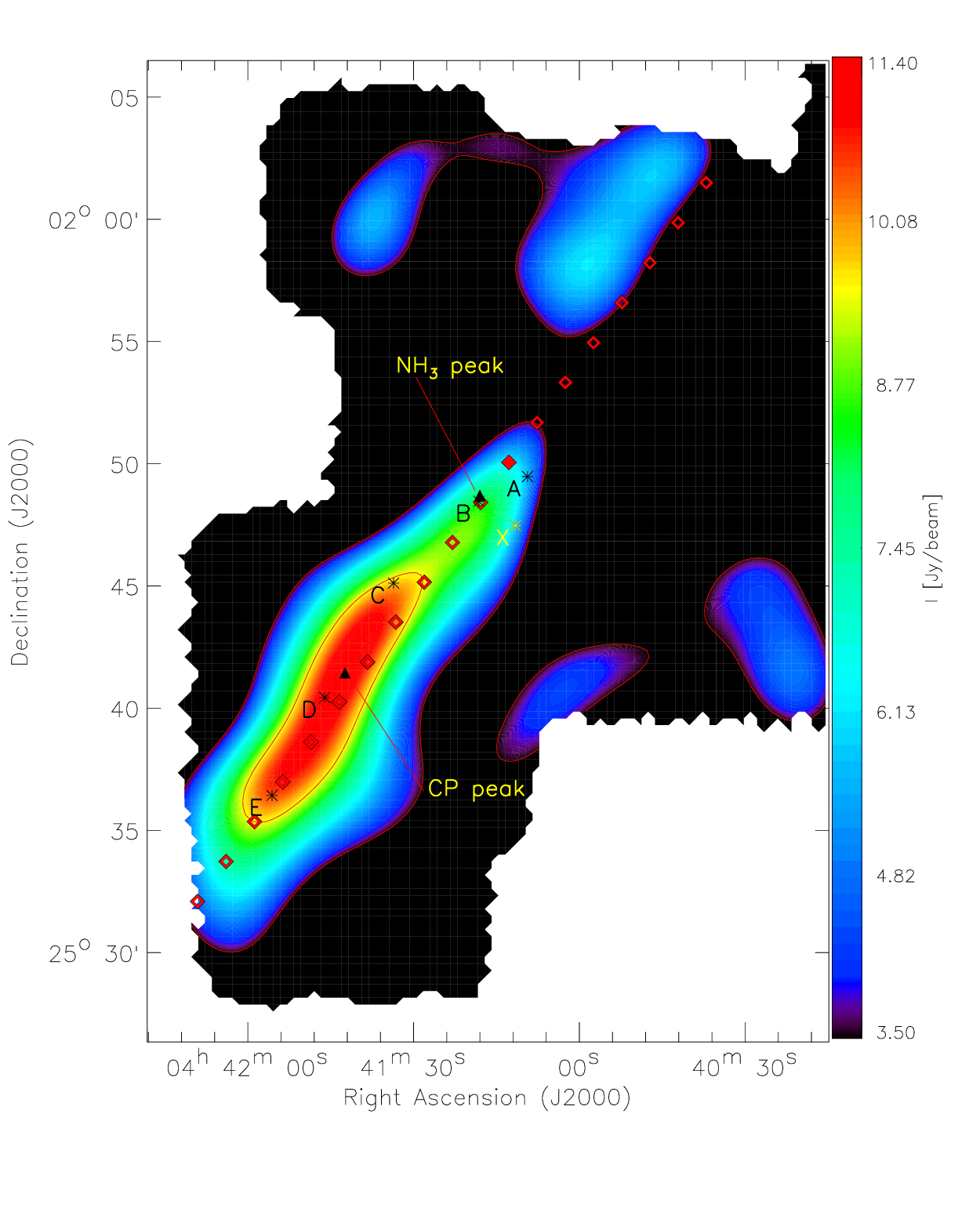}}
  \caption{\object{TMC-1} as seen by SCUBA at $850\ \rm{\mu m}$ (convolved to
  $30\arcsec$ in a and $228\arcsec$ in b). The
  map is reproduced from \cite{Nutter08}, who calls the filament 
  ``the Bull's Tail''.  The red diamonds represent the locations of our CH
   spectral observations (separated by $2\arcmin$), and the black
  triangles mark the locations of the cyanopolyyne and ammonia maxima.
  The letters A-E and X mark the regions of the six dense cores as per
  \cite{Hirahara92}, with the asterisks marking the locations of the
  CCS maxima corresponding to the six cores.  The two contour line
  levels correspond to intensities of 0.19 and 0.32 Jy/beam on the
  $30\arcsec$ resolution map and 3.5 and 9.6 Jy/beam on the
  $228\arcsec$ map.  The filled red diamond on the image is located at
  $\alpha_{\rm{J2000}}=04^{\rm{h}}41^{\rm{m}}12\fs7$,
  $\delta_{\rm{J2000}}=25\degr 50\arcmin 04\arcsec$, marking the
  origin of the CH observations performed by us.}
\end{figure*}

\subsection{\rm{CH} spectra}

The $\Lambda$-doubling line of CH in the ${}^2\Pi_{1/2},J=1/2$ ground
state ($\lambda = 9\ \rm{cm}$) was observed in August 2003 using the
100-m radio telescope, located in Effelsberg, Germany, belonging to the
Max-Planck-Institut f$\rm{\ddot{u}}$r Radioastronomie. The acquired
data includes observations of the main spectral line 1-1 and the
satellite lines 1-0 and 0-1. The frequency is $3335.481\ \rm{MHz}$ for
the main line, $3263.794\ \rm{MHz}$ and $3349.193\ \rm{MHz}$ for the
lower and upper satellites. The FWHM of the antenna at these
frequencies is approximately $3\arcmin.8$. During the observations the
system temperature varied approximately between $55\ \rm{K}$ and $60\
\rm{K}$.

The signal was fed into a 1024 channel autocorrelator which operated
as a 3-band system, with the main line occupying 512 channels with a
bandwidth of $391\ \rm{kHz}$. The satellite lines occupied 256
channels each with a bandwidth of $195\ \rm{kHz}$. With this setup the
frequency resolution of each band is $\Delta \nu = 763\ \rm{Hz}$,
translating into a velocity resolution of $\Delta v_{11} \approx 69\
\rm{m/s}$ for the main line. Calibration of the observations was done
using flux densities from \cite{OttCali} of the well-known radio
calibrators 3C84 and 3C123. The antenna temperature was observed to
not be dependent on elevation as expected at these wavelengths.

\begin{figure}
  \resizebox{\hsize}{!}{\includegraphics{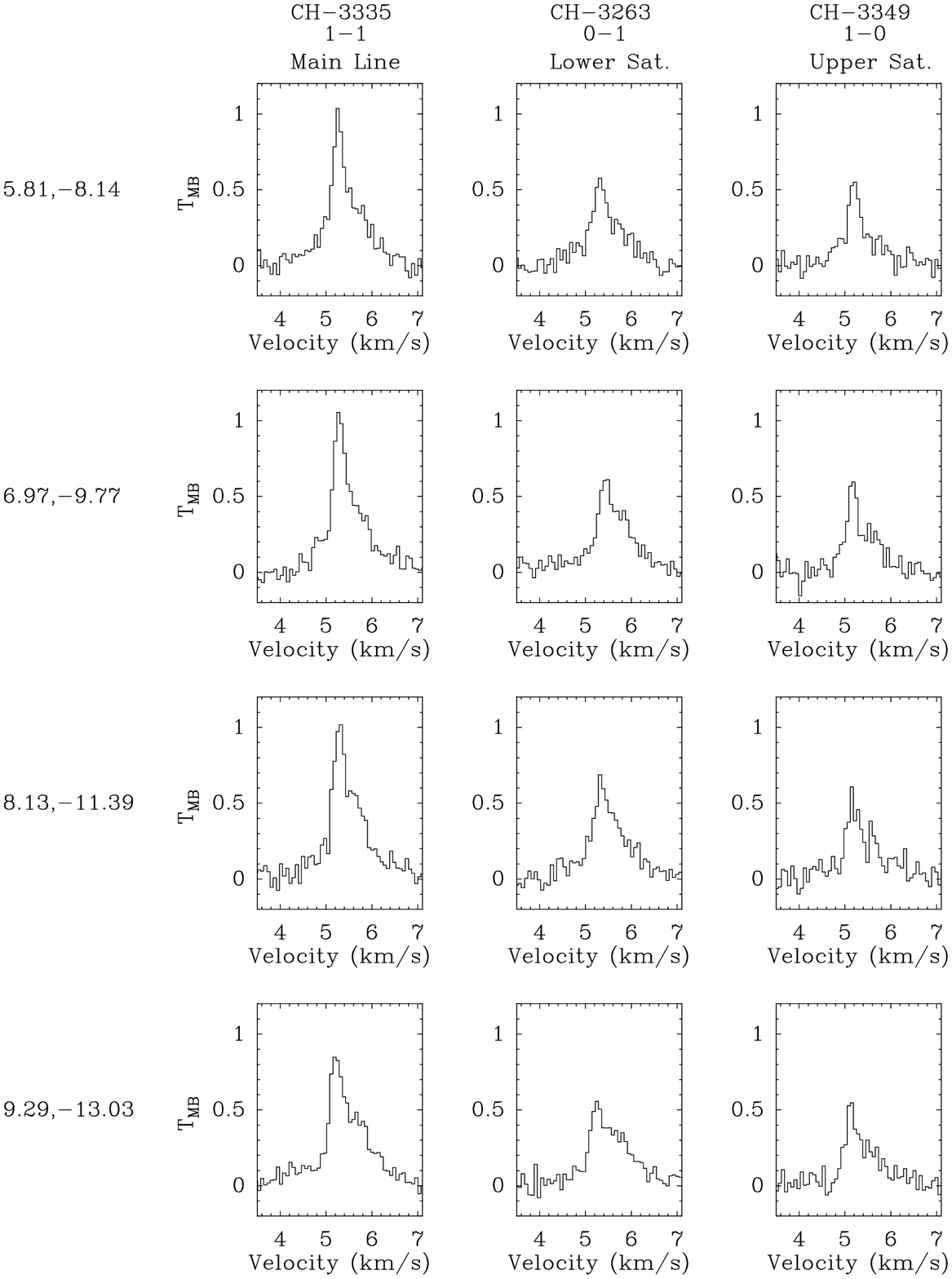}}
  \caption{Some of the reduced observations in the vicinity of the
  cyanopolyyne maximum for each observed CH spectral line. The numbers
  on the left indicate the right ascension and declination offsets in
  arcminutes from the center of the observations at 
  $\alpha_{\rm{B1950}}=04^{\rm{h}}38^{\rm{m}}08^{\rm{s}}.7$,
  $\delta_{\rm{B1950}}=25^{\rm{\circ}}44\rm{\arcmin}20\rm{\arcsec}$.}
  \label{fig:spectra}
\end{figure}

The observations were done using frequency switching. The origin of
the observations (the 0,0 offset) was set at
$\alpha_{\rm{J2000}}=04^{\rm{h}}41^{\rm{m}}12\fs7$,
$\delta_{\rm{J2000}}=25\degr 50\arcmin 04\arcsec$
($\alpha_{\rm{B1950}}=04^{\rm{h}}38^{\rm{m}}08\fs7$,
  $\delta_{\rm{B1950}}=25\degr 44\arcmin 20\arcsec$)
which corresponds to the $\rm{C^{18}O}$ maximum in the map of 
\cite{Langer95}. The observations followed a line through this point at a
$54^{\rm{\circ}}.5$ tilt relative to the B1950 declination axis. This line
passed over the cyanopolyyne maximum (\cite{Olano88}, located near our
position $\Delta x=+12\arcmin$) 
and the observation points
partially overlapped with the previous OH observations by
\cite{Harju00b}. A total of 22 different offsets were observed at the
3 wavelengths, bringing the total number of acquired spectra to
66. The offsets ranged from $\Delta \alpha = -8\arcmin.13$, $\Delta \delta =
+11\arcmin.39$ to $\Delta \alpha = +16\arcmin.26$, $\Delta \delta = -22\arcmin.8$
. See Figs. \ref{fig:SCUBAmap} and \ref{fig:SCUBAmap228} for a visual representation
and Table \ref{tab:coldens} for a list of the offsets. In addition to presenting the locations 
of the lines of sight as offsets along the right ascension and declination axes we 
will present them with $\Delta x$, which represents 
direct angular distance from the center of observations or (0,0). An example of reduced spectra 
is presented in Fig.~\ref{fig:spectra}. Due to human error, some of the initial
observations were done using a slightly erroneous rest
frequency. These observations had to be corrected manually back to the
right rest frequency during the reduction phase.

Furthermore, during our
analysis we noticed that the CH emission peaks at an anomalous velocity
of roughly 5.2 km/s. As no other observed molecules in TMC-1 are known
to peak so prominently at this velocity and along these lines of sight, we initially considered the 
possiblity of CH tracing an unknown gas component in the line of sight of TMC-1.
However, the fact that the obtained CH profiles are shifted with respect
to OH aroused a suspicion that the CH velocities are incorrect as
the two molecules are known to be well correlated in diffuse gas.
We are grateful to the referee for pointing this out.

After finding no explanation for the possible error in the spectral
headers we looked for previous CH observations for reference. The
surroundings of the CP peak in TMC-1 have been previously observed in
the CH main line by \cite{Magnani95} using the NRAO 43 m telescope,
and these spectra were kindly provided for us by Loris Magnani.
gaussian fits to two of the Magnani \& Onello positions closest to our observations
provided velocity differences of $\sim$0.5 km/s when compared to
similar gaussian fits to our own observations. The peak velocities
of the Magnani \& Onello positions were also similar to other
molecules observed in TMC-1. The beam size of the 43 m telescope 
is $9\arcmin$ at this frequency and the velocity resolution of the spectra is 0.44 km/s.

During the observing run we also did some test observations towards L1642
(MBM20), which has been previously observed in CH by \cite{Sandell81}
using the Onsala 25 m telescope. Our CH main line spectrum
towards the (0,0) of Sandell et al. has a peak velocity of $-0.22\pm 0.05$ km/s
whereas the Onsala spectrum peaks at a slightly positive velocity:
$+0.2\pm 0.1$ km/s.

\begin{figure}
  \resizebox{\hsize}{!}{\includegraphics{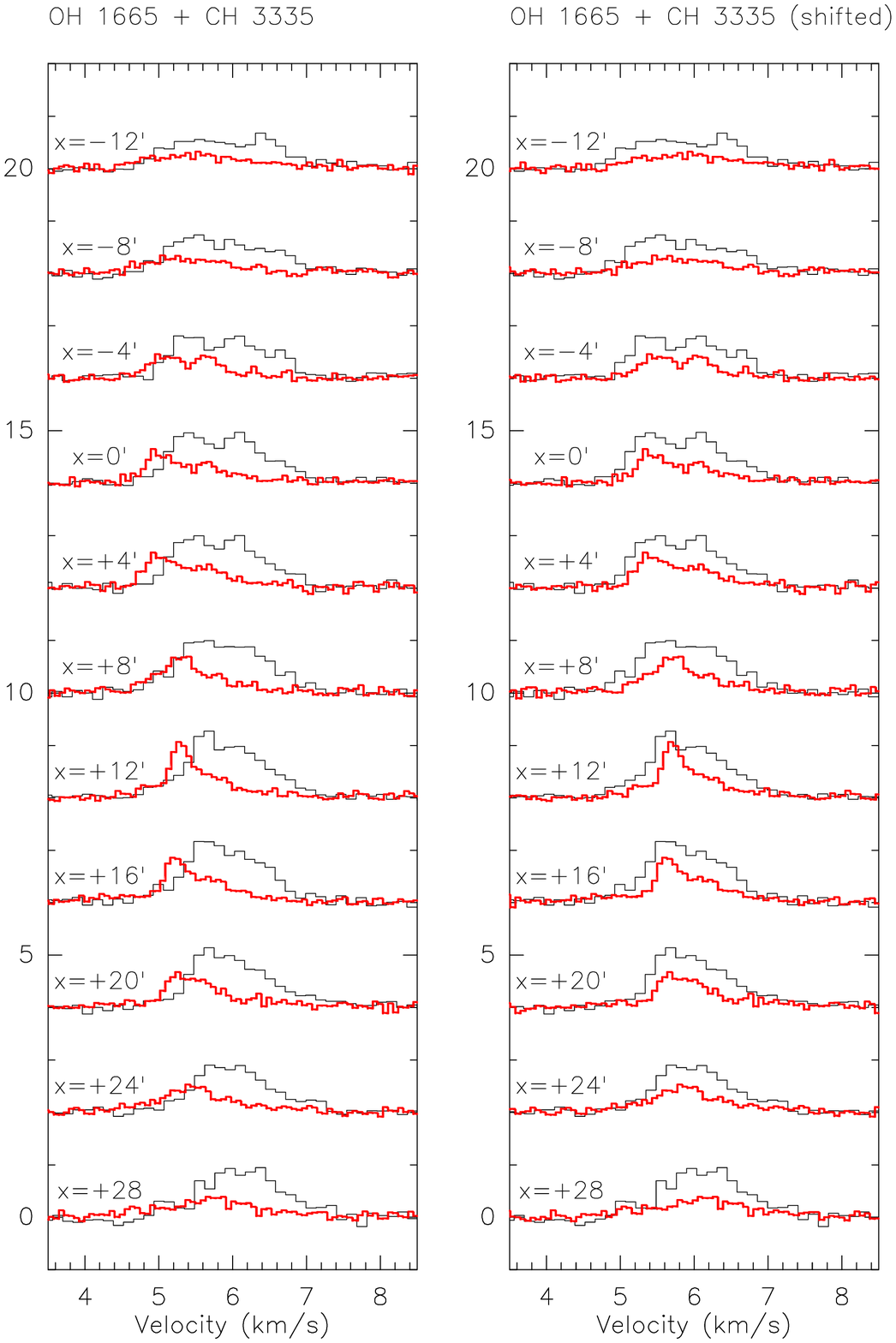}}
  \caption{An example of overlapping CH (thick red) and OH (thin black) spectra
  at several offsets before (left) and after (right) the CH spectra have been shifted to a velocity 6 channels
  higher than in our reduced data.}
  \label{fig:errorshift}
\end{figure}

After these comparisons we arrived to the conclusion that our CH lines
have a spurious velocity shift which corresponds to about six
autocorrelator channels. The effect of this shift is demonstrated in
Fig. \ref{fig:errorshift} which shows the CH and OH line profiles in various locations
using both the original spectra as they come from the telescope, and 
spectra shifted by 6 channels (to an about 0.4 km/s larger
velocity). According to the Effelsberg telescope team an error of this
magnitude can have occurred in the conversion of the raw telescope
data spectra to the Class reduction program format we use, in case an offset
due to the quantization of the synthesizer frequencies in the IF
section was not correctly registered.

\section{Results}

\subsection{Calculation \rm{CH} column densities}

Due to the very small Einstein A-coefficients of the
transitions, the $\Lambda$-doubling lines of
CH are optically thin. We can
furthermore assume that practically all CH molecules are in the ground
rotational state and that the $\Lambda$-doubling energy levels are
populated according to their statistical weights. The latter of these
assumptions is valid because the separations between the levels are
only $\sim 0.16$ K.

Absorption measurements have shown that for CH the ground state
$\Lambda$-doubling transitions are weak masers
\citep{RydBible,Hjalmarson77}. In the case of the main line, the
excitation temperature $T_{\rm{ex,11}}$ for CH is in the range $\sim
-60 \rightarrow -10$ K.

Armed with this knowledge and the integrated intensity of the CH main
line, $\int T_{\rm{MB}}dv\ [\rm{Kkms^{-1}}]$, the total column density
(in $\rm{cm^{-2}}$) of CH can be approximated with the formula
\begin{equation}
N({\rm CH}) = 2.82\, 10^{14} \frac{1}{1-T_{bg}/T_{ex,{\rm 11}}}
\int T_{\rm{MB}}(3335)dv
\label{eq:NCH_approx}
\end{equation}
where $T_{\rm MB}(3335)$ is the main-beam brightness temperature of
the main $\Lambda$-doubling component ($1-1$), $T_{\rm{bg}}$ is the
cosmic background temperature and $T_{\rm{ex},11}$ is the excitation
temperature of the $1-1$ transition.

In reality the assumption of statistical population distribution of
the $\Lambda$-doubling energy levels is not precisely valid, which can
be seen from the relative intensities of the three line components
(Fig.~\ref{fig:spectra}; both satellites should be two times weaker
than the main component).  The total column density can be given as a
function of the upper F=1 column density $N_{\rm{u},1}$ and the
excitation temperatures $T_{\rm{ex},11}$, $T_{\rm{ex},10}$ and
$T_{\rm{ex},01}$:
\begin{equation}
N({\rm CH}) = N_{{\rm u},1}\left(1+e^{\frac{T_{11}}{T_{{\rm ex},11}}}+
\frac{1}{3} e^{\frac{T_{10}}{T_{{\rm ex},10}}}+ \frac{1}{3}
e^{\frac{T_{11}}{T_{{\rm ex},11}}-\frac{T_{01}}{T_{{\rm ex},01}}}\right)
\label{eq:NCH}
\end{equation}
We have used the observed relative integrated intensities of the line
components, denoted by $K_{x,y}= \int T_{\rm{MB,x}}dv /\int
T_{\rm{MB,y}}dv$ to derive the excitation temperatures
$T_{\rm{ex},10}$ and $T_{\rm{ex},01}$ as functions
$T_{\rm{ex},11}$. The relative integrated intensities stay
approximately constant throughout the observed filament. We obtain
$K_{11,10}\approx 2.30$ and $K_{11,01}\approx 1.44$ by taking an
arithmetic mean of their respective area ratios from all the data.

A value for $T_{\rm{ex},11}$ must be chosen for us to be able to get
the CH column density. We choose $T_{\rm{ex},11}=-10\ \rm{K}$ which is
similar to the value derived by \cite{Hjalmarson77} for the dark cloud
LDN 1500 in front of 3C123.  This choice implies that
$T_{\rm{ex},10}\approx -28.8\ \rm{K}$ and $T_{\rm{ex},01}\approx -3.5\
\rm{K}$. Using the value of $T_{\rm{ex,11}}=-60\ \rm{K}$ from
\cite{GenzelEx} would result in an approximately $23\ \%$ increase in
the calculated CH column densities. Using equation \ref{eq:NCH_approx} for the column
density calculations would have produced only a $0.3\ \%$ difference
to these results.

\subsection{A closer look at the \rm{CH} spectra}
\label{sec:shift}

\begin{figure}
  \resizebox{\hsize}{!}{\includegraphics{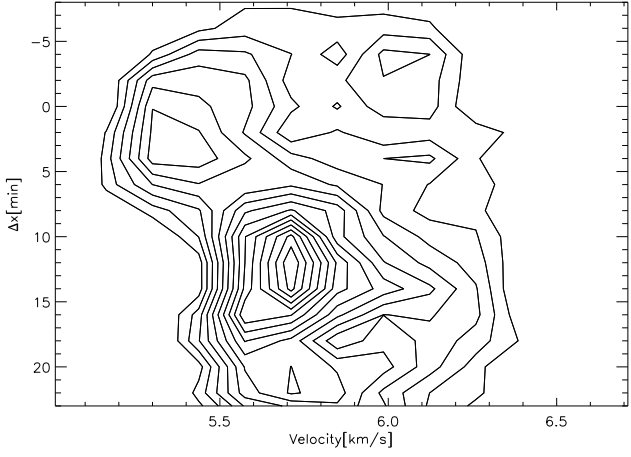}}
  \caption{Position-velocity diagram of the main CH spectral line. The lowest contour line corresponds to 0.27 K, increasing by about 0.05K per contour}
  \label{fig:vdmap}
\end{figure}

\begin{figure}
  \subfloat[][]{\resizebox{\hsize}{!}{\label{fig:vz3335}\includegraphics{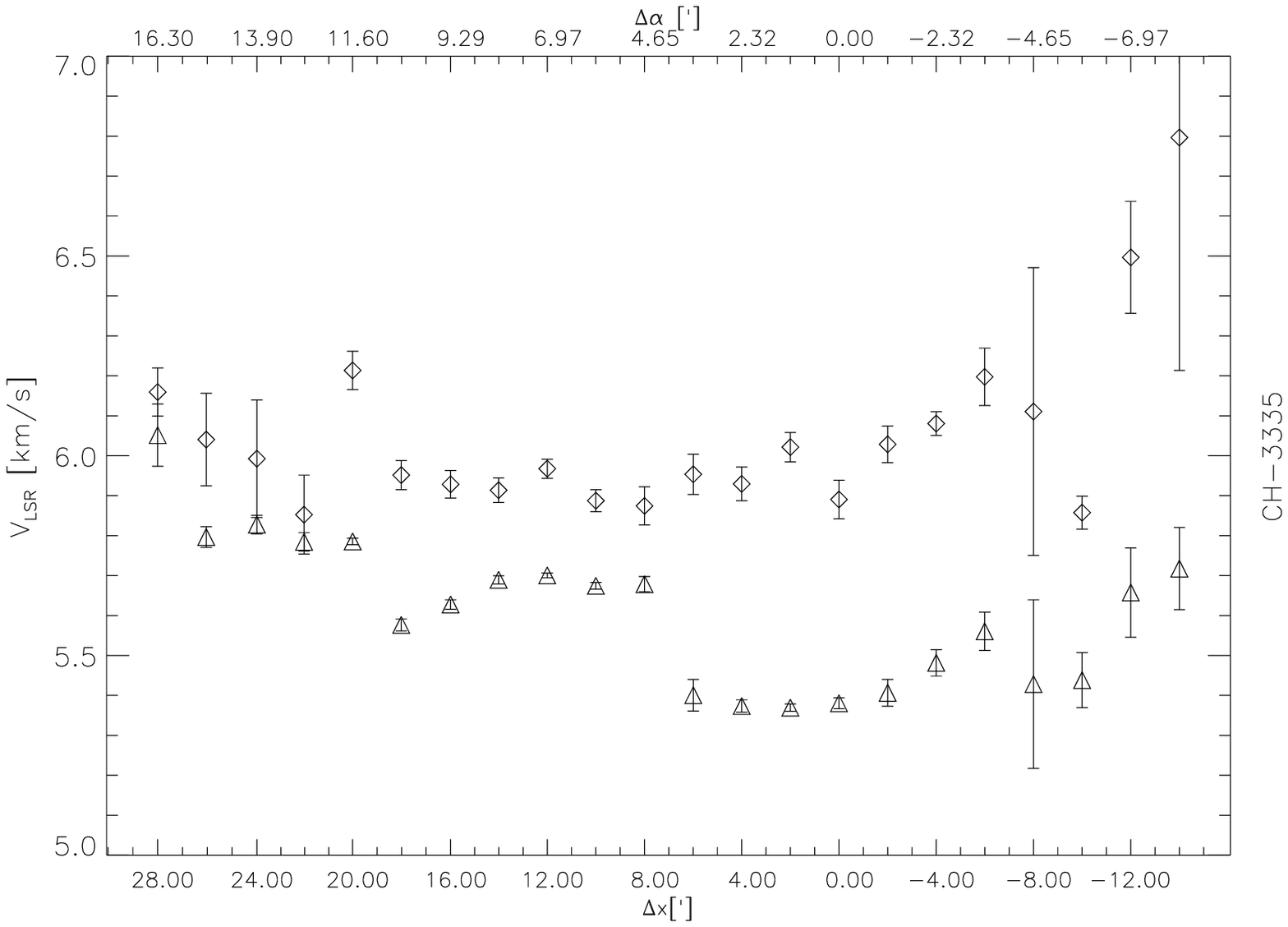}}}\\
  \subfloat[][]{\resizebox{\hsize}{!}{\label{fig:fwhm3335}\includegraphics{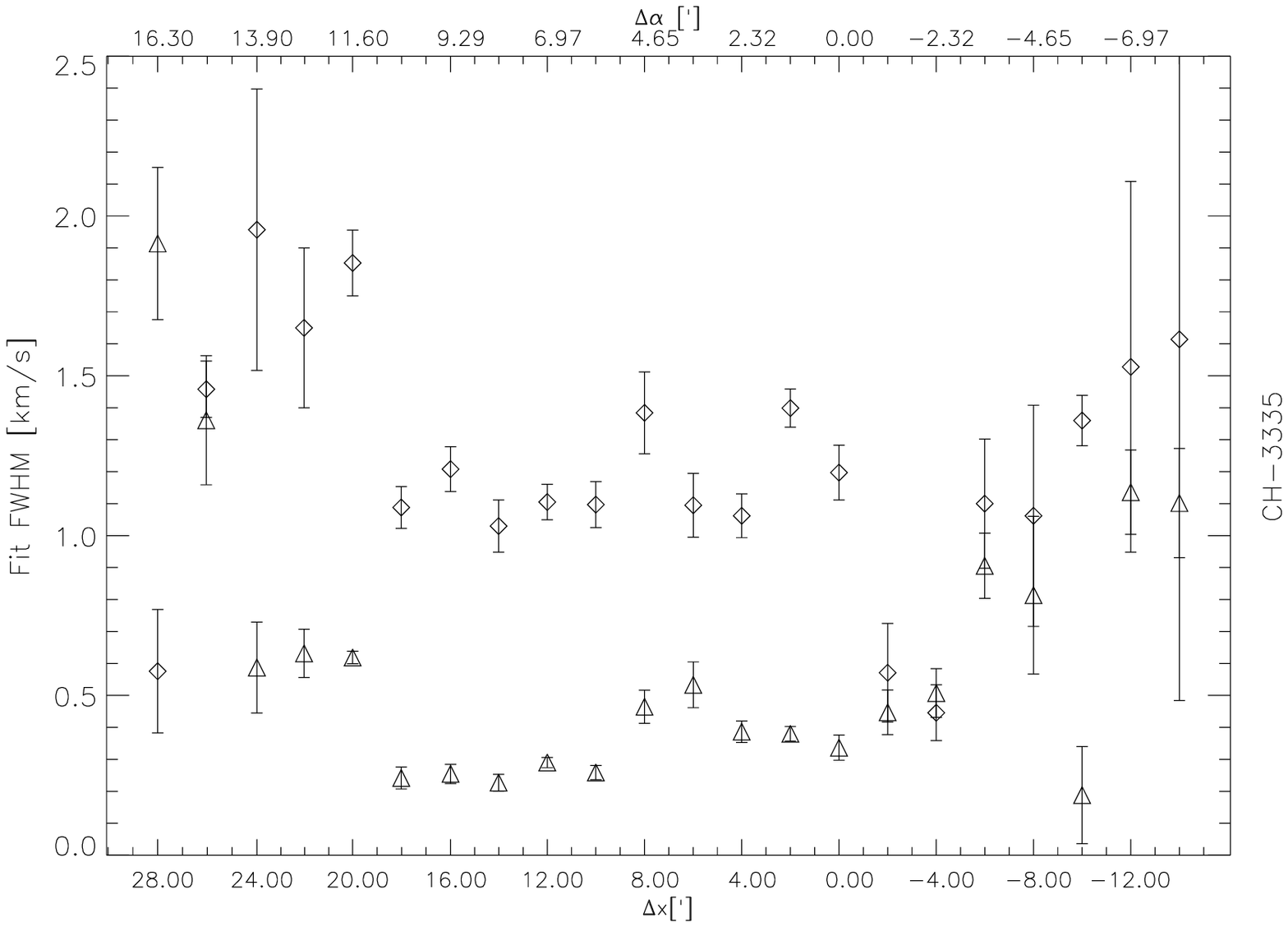}}}
  \caption{(a) Centroid positions and (b) widths of the two gaussian fits made
  to the main CH line at different offsets. The triangles represent
  the blueshifted component of the two fits.}
\end{figure}

The CH profiles are asymmetric in the sense that they peak close to
the lower boundary of the radial velocity range of the detectable
emission (see Fig.~\ref{fig:spectra}). It was noticed that at several
offsets it is possible to fit two gaussian profiles to the observed
spectra, especially for the main $\rm{CH}$ line: a narrow line
component peaking between 5.4 and 5.7 km\,s$^{-1}$, and a broader,
weaker component roughly between 5.8 and 6.4 km\,s$^{-1}$. In addition, a weak 
pair of wings can be discerned at some locations. The velocity distribution 
along the cloud axis is illustrated in Fig.~\ref{fig:vdmap} which shows
the position-velocity diagram, and in Figs.~\ref{fig:vz3335} and  
\ref{fig:fwhm3335} presenting the results of the gaussian fits. 

We see from Fig.~\ref{fig:vz3335} that the two gaussian components in
the CH spectra can be resolved between offsets $-4\arcmin \ldots +18\arcmin$ along
our axis of observation.  Looking at the points in the SCUBA maps we
see that the offset $\Delta x = +18\arcmin$ lies near the southeastern tip
of the filament, whereas $\Delta x = -4\arcmin$ is already somewhat off from
the bright part on its northwestern side.

Of the aforementioned two gaussian components the more narrow
component starts off at about 5.4 km/s and then suddenly jumps to
$V_{\rm{LSR}} \approx 5.7\ \rm{km/s}$ at $\Delta x = +8\arcmin$ and again to
about 5.8 km/s at $\Delta x = +20\arcmin$. The peak of the broader component
persists at LSR velocities roughly between 6.0 and 6.1 km/s.  As can
be seem from Fig.~\ref{fig:fwhm3335}, the low-velocity components also
seems to become significantly narrower at $\Delta x = +10\arcmin$, halving
its FWHM. This halved line width is equivalent to thermal broadening
(as calculated from dust temperatures in Table \ref{tab:coldens}) at
these locations, implying that the CH emission of that line is emitted
from an area of the cloud where no other line broadening effects,
e.g. turbulence, are present in any significant part.  
We note that the narrowing of masing lines is significant at 
large negative optical thicknesses, $-\tau>>1$ 
(e.g. Spitzer 1978, Ch. 3.3.), but for CH lines with
 $|\tau| < 0.1$  the effect is negligible. 

At other offsets the width of the sharper line can not be entirely accounted
for by thermal broadening and the wider line seems to have a
significant component of non-thermal broadening. Overall the
non-thermal width of the sharp line varies between about 0 m/s and 400
m/s while the non-thermal width of the wide line persists between 1
km/s and 1.4 km/s.


\begin{figure}
  \resizebox{\hsize}{!}{\includegraphics{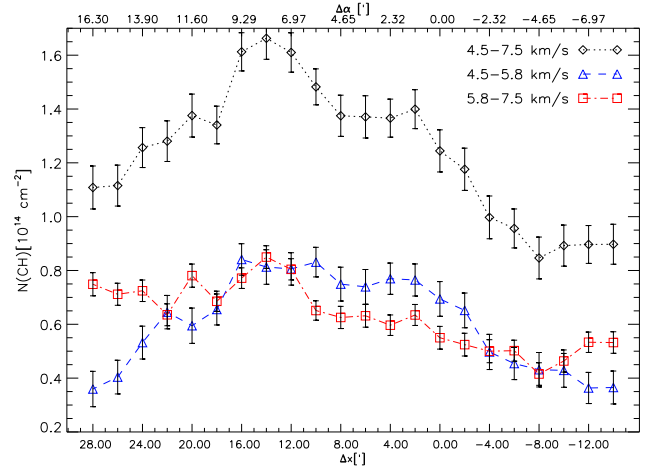}}
  \caption{Column densities for $\rm{CH}$ along the observed
    lines of sight in \object{TMC-1}. The black diamonds represent column
    densities calculated from the full velocity range and the
    blue triangles and red squares represent column densities calculated
    from velocity ranges corresponding to the narrow and wide 
    gaussian components respectively.}
  \label{fig:NCH}

\end{figure}

Due to the peculiar narrow blueshifted component, we decided to examine the 
calculated CH column densities not only in the full velocity range of 
detectable emission ($4.5-6.5$ km/s) but also with velocity ranges
roughly corresponding to the two separate line components. We calculated
$N$(CH) for the $V_{\rm LSR}$ range $4.5-5.4$ km/s for the narrow blueshifted
component and for the range $5.4-6.5$ km/s for the wider component. All three
behaviours of $N$(CH) along the filament are presented in Fig.~\ref{fig:NCH} and
the total column densities of CH are presented in number form in the fourth
column of Table \ref{tab:coldens}. 

In the figure we see that the narrow component seems to be the major contributing
factor to the slow increase of the total $N$(CH) starting at around $\Delta x = -6\arcmin$,
and its likewise gradual decrease at around $\Delta x = +20\arcmin$. The $N$(CH) contribution
of the broad component on the other hand seems to stay almost constant throughout the 
entire filament, increasing only very slightly as we go from the northwestern end of the
filament to the southeastern end. It does, however, have a distinctive ``bump'' in
the vicinity of the cyanopolyyne maximum and it seems to be the major contribution
 to the similar bump seen in the total $N$(CH) at that region.

 This bump is obviously also the maximum of the total 
column density in the entire region we measured and it is located approximately
at $\Delta x = 14\arcmin$, roughly between Hirahara's cores D and E.

\begin{table*}
\caption{Column densities of different molecules, visual extinctions and dust temperatures along the \object{TMC-1} filament. The velocity interval of the integrated main beam temperature used for calculating $N(\rm{CH})$ and $N(\rm{OH})$ is $4.5\ \rm{km/s}\ldots 7.5\ \rm{km/s}$. $\Delta \alpha$ and $\Delta \delta$ represent right ascension and declination offsets and $\Delta x$ represents direct angular distance from the center of the observations.}              
\label{tab:coldens}      
\centering                           
\begin{tabular}{c c c c c c c c c}        
\hline\hline                 
 $\Delta x\ \rm{[\arcmin]}$	&$\Delta\alpha\ \rm{[\arcmin]}$	&$\Delta\delta\ \rm{[\arcmin]}$	&$N(\rm{CH})\ \rm{[}10^{14}\ \rm{cm^{-2}]}$	& $N(\rm{OH})\ \rm{[}10^{15}\ \rm{cm^{-2}]}$	&  $A_{\rm V}\ \rm{[mag]}$	&$T_{\rm d}\ \rm{[K]}$   &$N(\rm{H_2})\ \rm{[}10^{21}\rm{cm^{-2}}\rm{]}$	\\
\hline                        
 $+28.00$			&$+16.26$			&$-22.80$			& $1.11 \pm 0.08$				& $2.27 \pm  0.12$				&  $4.82\pm 0.18$	        &-			&$4.53 \pm	0.17$					\\
 $+26.00$			&$+15.10$			&$-21.17$			& $1.12 \pm 0.08$				& -                    				&  $4.90\pm 0.19$	        &-			&$4.61 \pm	0.18$					\\
 $+24.00$			&$+13.94$			&$-19.54$			& $1.26 \pm 0.07$				& $2.12 \pm  0.07$				&  $5.12\pm 0.18$	        &-			&$4.81 \pm	0.17$					\\
 $+22.00$			&$+12.78$			&$-17.91$			& $1.28 \pm 0.08$				& -                    				&  $5.89\pm 0.24$	        &$12.61      \pm 0.37$	&$5.54 \pm	0.23$					\\
 $+20.00$			&$+11.61$			&$-16.28$			& $1.38 \pm 0.08$				& $2.50 \pm  0.11$				&  $7.55\pm 0.37$	        &$14.69	     \pm 0.38$	&$7.10 \pm	0.35$					\\
 $+18.00$			&$+10.45$			&$-14.65$			& $1.34 \pm 0.07$				& -                    				&  $8.46\pm 0.52$	        &$15.49	     \pm 0.37$	&$7.95 \pm	0.49$					\\
 $+16.00$			&$+9.29$			&$-13.03$			& $1.61 \pm 0.07$				& $2.74 \pm  0.08$				&  $9.74\pm 0.64$	        &$14.12	     \pm 0.30$	&$9.15 \pm	0.60$					\\
 $+14.00$			&$+8.13$			&$-11.40$			& $1.66 \pm 0.08$				& -                     			&  $12.20\pm 0.89$	        &$13.04      \pm 0.26$	&$11.47 \pm	0.84$					\\
 $+12.00$			&$+6.97$			&$-9.77$			& $1.61 \pm 0.07$				& $2.69 \pm  0.05$				&  $11.00\pm 0.82$	        &$12.87      \pm 0.26$	&$10.34 \pm	0.77$					\\
 $+10.00$			&$+5.81$			&$-8.14$			& $1.48 \pm 0.07$				& -                     			&  $12.51\pm 0.92$	        &$12.77      \pm 0.26$	&$11.75 \pm	0.87$					\\
 $+8.00$			&$+4.65$			&$-6.51$			& $1.37 \pm 0.08$				& $2.68 \pm  0.07$				&  $14.04\pm 1.05$	        &$12.14      \pm 0.25$	&$13.20 \pm	0.98$					\\
 $+6.00$			&$+3.48$			&$-4.88$			& $1.37 \pm 0.08$				& -                     			&  $27.01\pm 7.39$	        &$11.52      \pm 0.25$	&$13.11 \pm	0.55$					\\
 $+4.00$			&$+2.32$			&$-3.26$			& $1.37 \pm 0.07$				& $2.54 \pm  0.08$				&  $120.92\pm 73.83$	        &$11.24      \pm 0.24$	&$13.42 \pm	0.58$					\\
 $+2.00$			&$+1.16$			&$-1.63$			& $1.40 \pm 0.07$				& -                     			&  $10.48\pm 53.10$	        &$11.66      \pm 0.26$	&$12.05 \pm	0.53$					\\
 $0.00$				&$0.00$				&$0.00$				& $1.24 \pm 0.08$				& $2.54 \pm  0.06$				&  $11.58\pm 0.82$	        &$10.95      \pm 0.28$	&$10.89 \pm	0.78$					\\
 $-2.00$			&$-1.16$			&$+1.63$			& $1.18 \pm 0.08$				& -                     			&  $9.72\pm 0.62$	        &$8.73      \pm 0.28$	&$9.13 \pm	0.58$					\\
 $-4.00$			&$-2.32$			&$+3.26$			& $1.00 \pm 0.08$				& $2.18 \pm  0.08$				&  $9.96\pm 0.50$	        &$6.67      \pm 0.29$	&$9.36 \pm	0.47$					\\
 $-6.00$			&$-3.48$			&$+4.88$			& $0.96 \pm 0.07$				& -                     			&  $9.94\pm 0.47$	        &$5.88      \pm 0.28$	&$9.34 \pm	0.44$					\\
 $-8.00$			&$-4.65$			&$+6.51$			& $0.85 \pm 0.08$				& $2.04 \pm  0.08$				&  $8.97\pm 0.44$	        &$5.86      \pm 0.26$	&$8.43 \pm	0.42$					\\
 $-10.00$			&$-5.81$			&$+8.14$			& $0.89 \pm 0.08$				& -                     			&  $8.67\pm 0.38$	        &$5.71      \pm 0.27$	&$8.15 \pm	0.36$					\\
 $-12.00$			&$-6.97$			&$+9.77$			& $0.90 \pm 0.07$				& $1.89 \pm  0.07$				&  $7.99\pm 0.35$	        &$6.22      \pm 0.27$	&$7.51 \pm	0.33$					\\
 $-14.00$			&$-8.13$			&$+11.40$			& $0.90 \pm 0.07$				& - 						&  $6.89\pm 0.35$	        &$6.93      \pm 0.27$	&$6.48 \pm	0.33$					\\
\hline
\end{tabular}
\end{table*}

\subsection{$\rm{H_2}$ column densities}
\label{sec:NH2}

The NICER method \citep{Lombardi01} was applied to 2MASS data
\nocite{2MASS} of the filament region of \object{TMC-1} in order to acquire
visual extinction values for the points overlapping with our CH
observations, with the angular resolution element corresponding to the
FWHM available in the radio observations. Then, by applying the
$E(J-K)/N(\rm{H_2}) = 5.4\, 10^{21}\ \rm{cm^{-2} mag^{-1}}$
relation from \cite{Harjunpaa96} and \cite{Cardelli89}, it
would normally be possible to get column densities for molecular
hydrogen from the extinction data alone. However, as the dense cores
along the filament have high extinction values, acquiring reliable
values for $A_{\rm V}$ around them by using the NICER-method alone is
unreliable due to very low star densities close to and inside the
cores.

As an alternative means of obtaining $\rm{H_2}$ column densities we
used submillimetre dust continuum maps. The 850 and 450 $\mu$m SCUBA
maps of the \object{TMC-1} region from \cite{Nutter08} were kindly provided for
us by David Nutter. After convolving them to the FWHM of our 9 cm
observations (228") these were used to estimate $N$($\rm{H_2}$) in the
positions observed in CH.
The scan-mapping mode used in the SCUBA
observations involves chopping by a maximum throw of 68${\rm \arcsec}$. The
continuum maps are not sensitive to spatial scales larger than
that. In order to compare the dust maps with the extinction and CH
line data, we need to convolve the 850 and 450 mum SCUBA maps to the
FWHM of 228${\rm \arcsec}$ of our 9 cm observations. We believe that in the vicinity
of the TMC-1 ridge where the dust emission is likely to be dominated
by compact cores this convolution results in reasonable "smoothed"
flux densities. Furthermore, we attempt to correct for the arbitrary background
level of the SCUBA maps by correlating the intensities with the extinction
values as desribed below.
The 850 $\mu$m SCUBA maps around the Bull's
Tail smoothed to 30${\rm \arcsec}$ and 228${\rm \arcsec}$ resolutions are shown in
Figs. \ref{fig:SCUBAmap} and \ref{fig:SCUBAmap228}, respectively.

The H$_2$ column density is related to the intensity
of the dust emission by
\begin{equation}
N({\rm H_2}) = \frac{I_{\rm \nu}}{B_{\rm \nu} (T_d) \cdot 2.8 m_H \cdot \kappa_{\rm \nu} \cdot R_d},
\label{eq:NH2}
\end{equation}
where $B_{\rm \nu}(T_{\rm d})$ is the Planck function, $T_{\rm d}$ is the
temperature of the dust, $m_{\rm H}$ is the mass of a hydrogen atom,
$\kappa_{\rm \nu}$ is the dust cross-section per unit mass of dust and
$R_d$ is the dust-to-gas ratio. For $\kappa_{\rm \nu}$ and $R_d$ we used the
values $0.13\,{\rm cm^2 g^{-1}}$ (at 850 $\mu$m) and 1/100, respectively.

As bolometer maps usually do, the SCUBA maps contain negative
artefacts, i.e. hollows, around bright emission regions.  This causes
an uncertainty to the absolute level of the intensity $I_{\rm \nu}$, and
suggests a need for performing a bias correction. We combined the
NICER-generated extinction map with the SCUBA maps to examine the
relation between extinction and submm intensity much in the same way
as was done by \cite{Bianchi03}. This examination can provide us with
two things: A bias correction for the SCUBA maps, and the average
temperature of the dust along the filament.

A linear fit of the form
\begin{equation}
I^{\rm{obs}}_{\rm \nu} = \delta_{\rm \nu} + \gamma_{\rm \nu} A_{\rm V}.
\label{eq:IAfit}
\end{equation}
was performed to the pixel by pixel correlations between the observed
intensities at 850 or 450 $\mu$m and the visual extinction. The fit
was done using pixels with $A_{\rm V} < 12$, because above this level
the extinctions are highly uncertain (see above), and because the dust
temperature, $T_{\rm d}$, is expected to decrease towards the most
obscure regions, causing a curvature in the correlation
\citep[see][]{Bianchi03}.

The coefficient $\delta_{\rm \nu}$ should be zero in a properly biased map. The
intensity of the dust emission is given by
\begin{equation}
I^{\rm{dust}}_{\rm \nu} = B_{\rm \nu} (T_{\rm d}) (1-e^{-\tau_{\rm \nu}}) \approx B_{\rm \nu}
(T_{\rm d}) \tau_{\rm \nu} \; ,
\end{equation}
where the optical thickness, $\tau_{\rm \nu}$, of the source of emission can
be written as $\tau_{\rm \nu} = \Sigma \kappa_{\rm \nu} R_{\rm d}$, where $\Sigma$
is the surface density of the cloud. It can be assumed that
subtracting the fitting parameter $\delta_{\rm \nu}$ from the 850 and 450
$\mu$m intensity maps should result in a better-than-nothing
correction to the bias.

The dust temperature was obtained from the ratio of the $\gamma$
coefficients at the two frequencies:
\begin{equation}
\frac{\gamma_{\rm \nu 1}}{\gamma_{\rm \nu 2}} =
\frac{\kappa_{\rm \nu 1}}{\kappa_{\rm \nu 2}} \frac{B_{\rm \nu 1}(T_d)}{B_{\rm \nu 2}(T_d)} =
\left(\frac{\nu_1}{\nu_2}\right)^{\beta} \frac{B_{\rm \nu 1}(T_d)}{B_{\rm \nu 2}(T_d)}
\; .
\label{eq:TdA}
\end{equation}
In the latter form, we have made use of the assumption that at
(sub)millimetre wavelengths $\kappa_{\rm \nu} \propto \nu^\beta$, where the
emissivity index $\beta = 1.95$ between 850 and 450 $\mu$m \citep[][]{Ossenkopf94}.

\begin{figure}
  \subfloat[][]{\resizebox{\hsize}{!}{\label{fig:masked850}\includegraphics{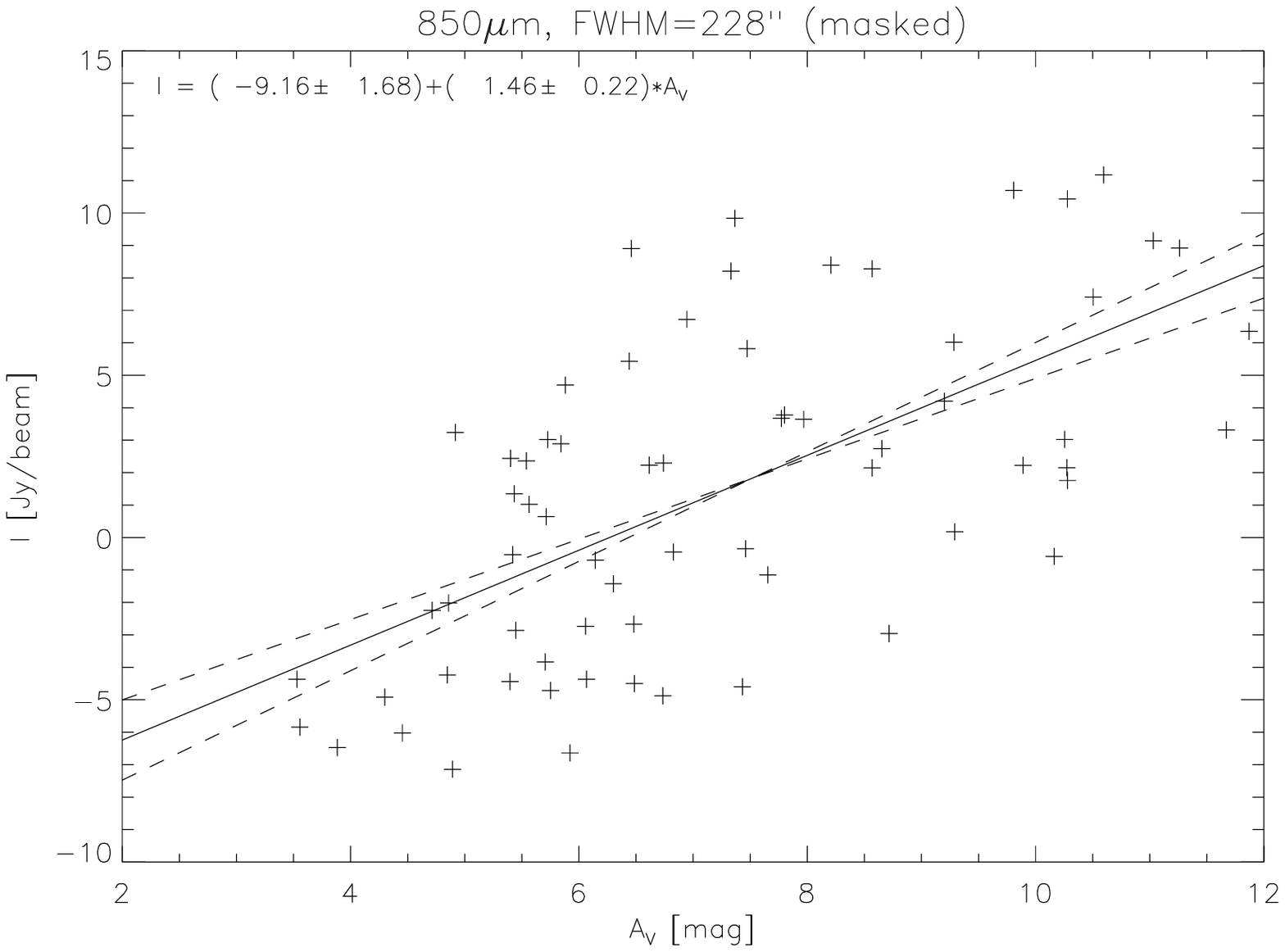}}}\\
  \subfloat[][]{\resizebox{\hsize}{!}{\label{fig:masked450}\includegraphics{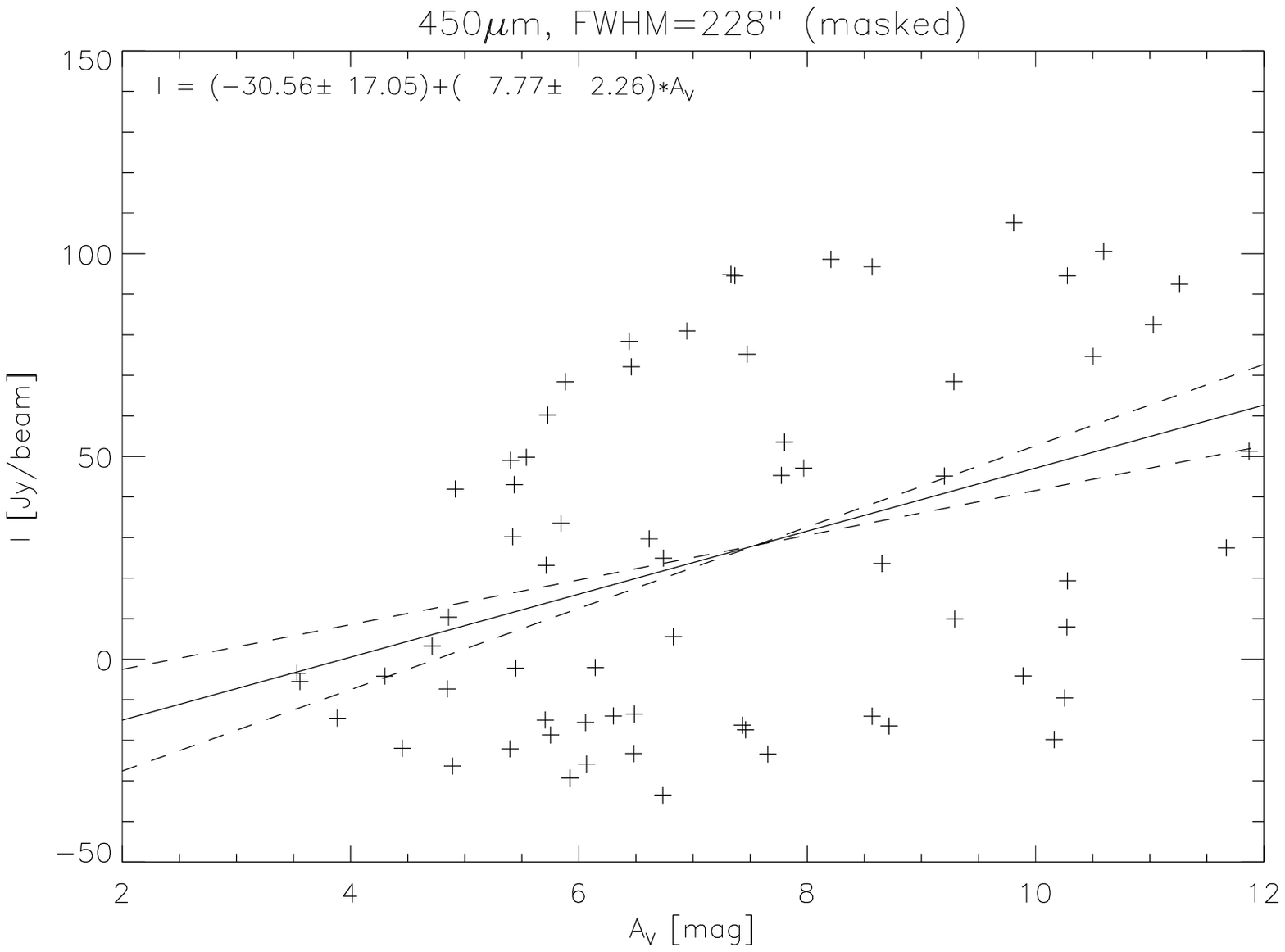}}}
  \caption{A linear fit to the visual extinctions and (a) $850\
  \rm{\mu m}$ or (b) $450\ \rm{\mu m}$ intensities around the filament
  region of \object{TMC-1}. The resolution of the intensities and the
  extinctions both correspond to a beam FWHM of $228\arcsec$.}
\end{figure}

\begin{figure}
  \resizebox{\hsize}{!}{\includegraphics{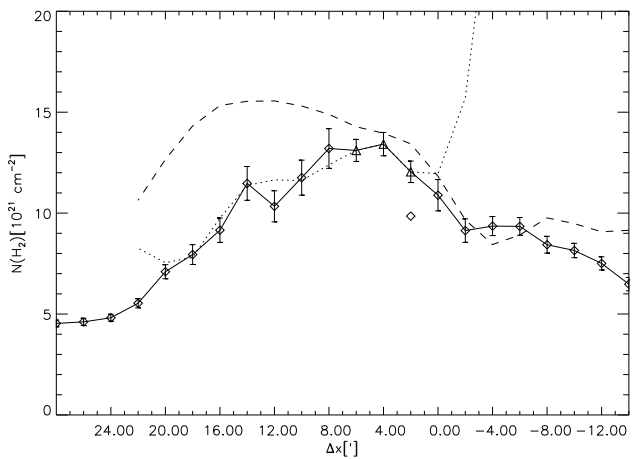}}
  \caption{Column densities for $\rm{H_2}$ along the \object{TMC-1}
  filament. The dashed line represents molecular hydrogen column
  densities calculated with Eq. \ref{eq:NH2} while assuming a constant
  dust temperature of $T_d \approx 11.0\ \rm{K}$. The diamonds
  represent $N$($\rm{H_2}$) values derived from a NICER extinction map
  as described in the beginning of section \ref{sec:NH2}. The triangles
  and the dotted line represent $N$(H$_2$) values calculated by using
  variable dust temperatures obtained from Eq. \ref{eq:TdB}.}
  \label{fig:NH2}
\end{figure}

As the dust temperature, and possibly also the ratio dust opacities in
the submm and in the visual, $\kappa_{\rm \nu}/\kappa_{\rm V}$, can be
expected to be different in the filament when compared to several
other parts of the cloud, the examination of the linear relation
between observed intensities and extinctions were done only on pixels
in a rectangular area passing along the filament with a width of about
$400\arcsec$. Linear fits were performed in these areas for comparing both
the $450\ \rm{\mu m}$ and $850\ \rm{\mu m}$ maps to the visual
extinctions as calculated with the NICER method. These comparisons and
the linear fits performed on them are presented in
Figs. \ref{fig:masked850} and \ref{fig:masked450}. With the help of
these fits we corrected the bias on the intensity maps, and calculated
the average dust temperature of the examined area to be $T_d = (11.0
\pm 2.7)\ \rm{K}$. After this, with the help of Eq. \ref{eq:NH2}, it
was possible to use the $850\ \rm{\mu m}$ (because it seemed more
accurate) SCUBA map to calculate the $\rm{H_2}$ column densities on
the points overlapping with our $\rm{CH}$ observations (three points
in the southeast were not covered by the SCUBA maps). The resulting
$N(\rm{H_2})$ estimates are plotted in Fig.~\ref{fig:NH2} as a dashed
line.

The column densities calculated in this fashion agree rather well with
those estimated from the $A_{\rm V}$ map on the northwestern side of
the map ($\Delta x \leq 0^\prime$, see Fig.~\ref{fig:NH2}), whereas
the agreement is not good in the southeast. However, it is conceivable
that the $A_{\rm V}$ map reflects the true H$_2$ column density
distribution except for the regions of the highest obscuration, in
particular it should do so at both ends of the strip observed in CH.

The discrepancy between the $N({\rm H_2})$ estimates from the 850
$\mu$m emission presented above and those from $A_{\rm V}$ can be
caused by the fact that assumption of a constant dust temperature is
not valid.  We examined therefore the possible variations of the dust
temperature along the filament by using the 'bias-corrected' 450
$\mu$m and 850 $\mu$m intensities in the formula
\begin{equation}
\frac{I^*_{\rm \nu_2}}{I^*_{\rm \nu_1}} \left(\frac{\nu_{1}}{\nu_{2}}\right)^{3.0+\beta}= \frac{e^{h\nu_{1} /kT_d}-1}{e^{h\nu_{2}/kT_d}-1}.
\label{eq:TdB}
\end{equation}
This method can be applied with a reasonable accuracy to the positions
within the dense filament where both 450 and 850 $\mu$m signals are
strong.  The resulting dust temperatures and H$_2$ column densities
are presented in Table \ref{tab:coldens} and as a dotted line in
Fig.~\ref{fig:NH2}. It is worth noting that the that, in the envelope,
the dust temperature calculated in this fashion is similar to the $T_{\rm d} = 12\ {\rm K}$
derived by Nutter et al. for the shoulder component. This is also quite
close to kinetic gas temperatures observed in the cloud by e.g. \cite{Pratap97}.

Since the extinction method seems reliable everywhere except at three
points ($\Delta x=2\arcmin\ldots 6\arcmin$) around the calculated extinction value
maximum and because the dust temperature seems to be variable along
the filament, we have opted to use the extinction-derived $N({\rm H_2})$ values
for all except the three aforementioned points. For these three points
we use instead the $N({\rm H_2})$  values calculated by using dust temperatures
acquired from Eq. \ref{eq:TdB}. These column densities are listed in
the sixth column of Table \ref{tab:coldens}.

\subsection{\rm{CH} abundance }

\begin{figure}
  \resizebox{\hsize}{!}{\includegraphics{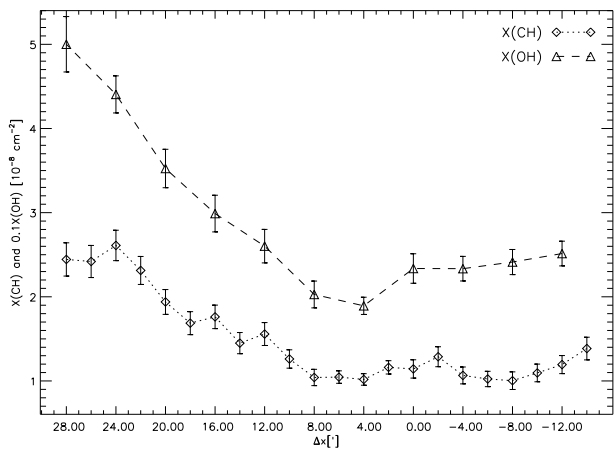}}
  \caption{The abundances of CH and OH along the observed lines of sight
    in \object{TMC-1}.}
  \label{fig:CHrat}
\end{figure}

When comparing the column densities of $\rm{H_2}$ and $\rm{CH}$ in an
x,y-plot, the slope of a linear fit performed on the data points
should tell the average abundance of CH in the observed area. This
approach is however not very interesting if the two column densities
do not correlate very well. No strong correlation ($\rho \approx
0.53$) between the two column densities was noticed, suggesting that
the fractional abundance of CH is not very constant throughout the region. The
abundance of CH, or $X({\rm CH})=N({\rm CH})/N({\rm H_2})$, along the
observed axis is presented in Fig.~\ref{fig:CHrat}. This figure shows, beginning from
$\Delta x=+8\arcmin$,
a steady rise in $X(\rm{CH})$ from about $1.0\, 10^{-8}$ up to
$\sim 2.2\, 10^{-8}$ towards the southeastern end of the
filament. Our result at the cyanopolyyne peak matches the value
quoted in \cite{Ohishi92}. Examining the abundance
of CH calculated by using $N$(CH) from the narrow and wide
line components is not very useful because we
can not evaluate $N$(H$_2$) from the two different velocity
intervals separately and thus we refrained from doing so.
\subsection{Correlation with $\rm{OH}$}

\begin{figure}
  \resizebox{\hsize}{!}{\includegraphics{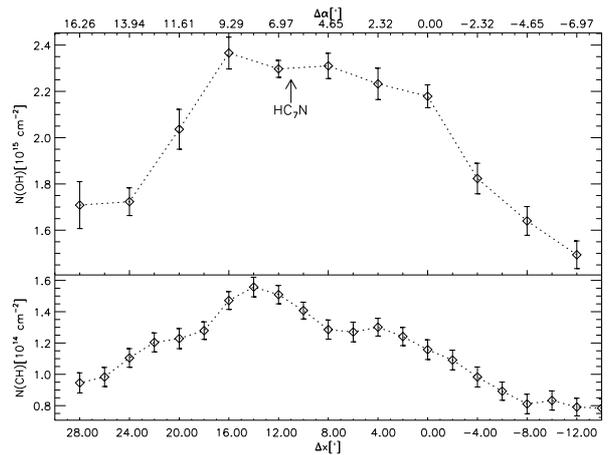}}
  \caption{Column densities for $\rm{OH}$ and CH along the \object{TMC-1}
  filament. The arrows point out the approximate location of the
  cyanopolyyne maximum as per \cite{Olano88}.}
  \label{fig:NOH}
\end{figure}

We had access to previous OH spectral line data acquired by
\cite{Harju00b} with partially overlapping points to our CH
observations and used the formula from the same paper to acquire the
column densities for OH:
\begin{equation}
N({\rm OH})=4.04\, 10^{14} \frac{1}{1-T_{\rm bg}/T_{\rm ex}}\cdot 
\frac{1}{\eta_{\rm C}}\int T_{\rm MB}(1665)dv
\label{eq:NOH}
\end{equation}

with $[\rm{Kkms^{-1}}]$ as the unit for the line area and
$[\rm{cm^{-2}}]$ as the unit for the resulting column density. The
excitation temperature and main beam efficiency used for this was
$T_{\rm{ex}}=10\ \rm{K}$ and $\eta_{\rm{C}}=30\ \%$. The velocity
interval ($4.5\ \rm{km/s}\ldots 6.5\ \rm{km/s}$) of the integrated
line area was chosen to be identical to the one used for the CH column
density calculations. The resulting column densities are listed in the
fifth column of Table \ref{tab:coldens} and they are plotted in Fig. \ref{fig:NOH}.

\begin{figure}
  \resizebox{\hsize}{!}{\includegraphics{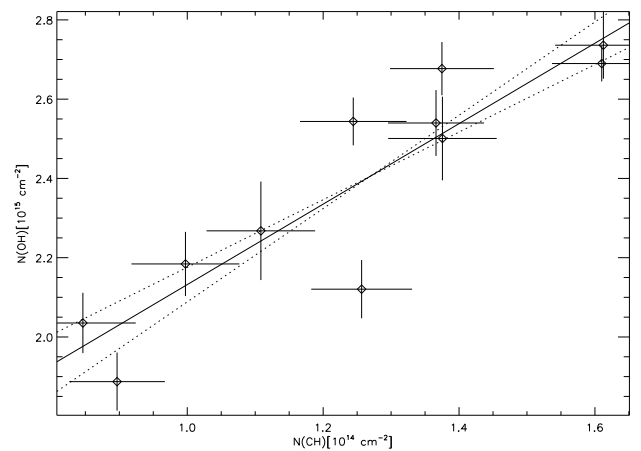}}
  \caption{Correlation plot of the column densities of CH versus OH in
  spots where observations of them overlap.}
  \label{fig:CHvOH}
\end{figure}

As can be seen from Table \ref{tab:coldens}, there are twice as many CH
observations as there are OH observations along the strip we are
investigating. To make use of all the CH observations when comparing
them to OH, we have used a simple but sufficient averaging scheme
where, on each overlapping CH and OH observation, we take a weighted
average of CH and its adjacent point on each side and give the
adjacent points a weight of one half.
The overlapping observations are plotted to Fig.~\ref{fig:CHvOH} along
with a linear fit made between the points. A strong correlation
($R_{\rm{CH,OH}}\approx 0.90$) is noticed between the column densities
and the column densities follow the relation

\begin{equation}
  N({\rm OH})[{\rm cm^{-2}}]=(11.2\pm 2.1)\, 10^{14} + 
(10.1\pm 1.6)\cdot N({\rm CH}) \; .
\end{equation}

The fractional abundance of OH along the \object{TMC-1}
ridge is plotted
together with that of CH in Fig.~\ref{fig:CHrat}. 
As the CH column density lies
in the range $\sim 0.9-1.6 \, 10^{14}$ cm$^{-2}$ the derived
OH/CH abundance ratio along the ridge, $N$(OH)/$N$(CH)$\sim 16-20$, is
clearly larger than that measured recently in diffuse interstellar gas
by \cite{Weselak09}: $N$(OH)/$N$(CH)$\sim 2.6$, which is close to the
atomic O/C ratio in the local ISM \citep[e.g.][]{InterSChem,Wilson94}.

\section{Discussion}

\subsection{The chemistry of \rm{CH} and its relation to \rm{OH}}
\label{sec:chemistry}
Pure gas-phase chemistry models \citep[e.g.][]{Viala86} predict that
the CH abundance, $X({\rm CH})$, decreases from $\sim 10^{-8}$ at
the outer boundaries of molecular clouds ($A_{\rm V} \leq 1$) to $<
10^{-9}$ towards the cloud interiors. This change should imply
saturation in the correlation of $N({\rm CH})$ 
with $N({\rm H_2})$ towards
dense molecular clouds, and indeed this kind of tendency can be traced
in observational results \citep{Hjalmarson77,Mattila86,Qin10}.

The production of CH depends on the availability of C$^+$ or C. C$^+$
is the dominant form of carbon
in the outer parts of a cloud exposed
to UV radiation. In the inner parts, C$^+$ formation is hindered
by the lack of UV photons and if formed is rapidly neutralized or
incorporated into neutral molecular species via ion-molecule and
dissociative recombination reactions. When the elemental O/C ratio is larger
than unity, most carbon is consumed by the formation of gas-phase CO, 
which eventually freezes onto grain surfaces 
at sufficiently low temperatures, $\lesssim$ 20~K.

The principal source of CH in dark clouds is likely to be the
dissociative recombination of hydrocarbon ions, in particular CH$_3^+$
and CH$_5^+$, e.g.:
\begin{equation}
{\rm CH}_3^+ + e^- \rightarrow {\rm CH} + {\rm H_2} \; .
\label{eq:1}
\end{equation}
The sequence leading to these ions is initiated either by the radiative
association of C$^+$ and H$_2$:
\begin{equation}
{\rm C}^+ + {\rm H_2} \rightarrow {\rm CH_2^+} + h\nu \; ,
\label{eq:2}
\end{equation}
or by the proton exchange reaction between neutral C and H$_3^+$:
\begin{equation}
{\rm C} + {\rm H_3^+} \rightarrow {\rm CH^+} + {\rm H_2}
\label{eq:3}
\end{equation}
\citep[e.g.][and references therein]{Black75}. The sequence continues
to larger hydrocarbon ions through reactions with H$_2$.
CH is thus expected to correlate with C$^+$ and \ion{C}{I}, and because
these species are necessary to the production of complex carbon
compounds, it should correlate also with heavier hydrocarbons and
cyanopolyynes. Furthermore, a close relationship between CH and  
long carbon chains or cyclic molecules may originate in direct CH insertion  
into an unsaturated hydrocarbon (followed by H elimination) which is likely 
to proceed with a small or no barrier \citep[e.g.][and references therein]{Soorkia10}.

The destruction of CH is thought to occur primarily via the following
neutral exchange reactions:
\begin{equation}
\begin{array}{ccc}
{\rm CH} + {\rm O} &\rightarrow& {\rm CO} + {\rm H} \\
{\rm CH} + {\rm N} &\rightarrow& {\rm CN} + {\rm H} \;.
\label{eq:4}
\end{array}
\end{equation}

In contrast with pure gas-phase chemistry, models including accretion
of molecules onto grain surfaces predict a smooth distribution for the
CH abundance in dense clouds. \cite{Hartquist89} and 
\cite{Rawlings92} have pointed out that the CH abundance may remain
almost unchanged in the dense interiors of dark clouds where heavy
molecules are likely to be depleted. As long as some gas-phase
CO is available,
carbon ions are supplied by reaction with He$^+$ 
(which is continually
produced by cosmic rays):
\begin{equation}
{\rm He^+} + {\rm CO} \rightarrow {\rm C^+} + {\rm O} + {\rm He} \;.
\label{eq:5}
\end{equation}
The reduction of CO is partly compensated by the diminishing
destruction of CH by O or N. Also an increase in the H$_3^+$ abundance
associated with the depletion can have a balancing effect on
the CH
abundance. Furthermore, the adsorption energy of CH is relatively low and
it is efficiently returned to the gas-phase via thermal desorption
\citep{Aikawa97}.
\\

\vspace{2mm}

In dark clouds, the reactions leading to the formation and destruction
of OH are probably analogous to those for CH. It is thus expected that
OH is mainly produced by dissociative recombination of H$_3$O$^+$ and
destroyed in neutral exchange reactions with C, N, and O atoms. In
order to identify the most important reactions and to study temporal
changes in the OH/CH abundance ratio, we ran some chemical models 
which, although relatively simple, are 
appropriate for dark clouds such as \object{TMC-1}.

We modeled the chemistry of \object{TMC-1} by treating the source as a
homogeneous, isotropic cloud with the constant physical parameters,
$n({\rm H_2}) = 10^{4} \, {\rm cm}^{-3}$, $T = 10$ K (for both gas and dust)
and $A_{\rm V} = 10$ mag, $\sigma_{\rm g} \approx 0.03\ \rm{\mu m}$,
using a cosmic-ray ionisation rate of $1.3 \, 10^{-17}\,{\rm
  s}^{-1}$ and performing a pseudo-time-dependent calculation in which
the chemistry evolves from initial abundances that are atomic,
excepting molecular hydrogen. We adopt the low-metallicity oxygen-rich
elemental abundances from \cite{Graedel82} and \cite{Lee98}
as listed in Table 8 of \cite{UMIST}. Thus, our initial
elemental abundances with respect to H nuclei number density for C and
O are $7.3\, 10^{-5}$ and $1.76\, 10^{-4}$, respectively, giving a C/O
ratio of ~ 0.4. Note, under dark cloud conditions, hydrogen exists
primarily in molecular form so that the H nuclei number density is
twice that of molecular hydrogen. Our gas-phase chemical network is
the latest release of the UMIST Database for Astrochemistry or
\textcolor{red}{``}Rate06" (http://www.udfa.net) as described in \cite{UMIST}.
We ran one model using gas-phase chemistry only and another allowing
for gas-grain interactions, namely, the accretion of gas-phase species
onto dust grains and the thermal desorption of species from the grain
surfaces. For our second model, we assume the dust grains are
spherical particles with a radius of $10^{-5}$ cm and that the dust is
well mixed with the gas possessing a constant fractional abundance of
$2.2\, 10^{-12}$ with respect to H nuclei density. Our accretion and
thermal desorption rates were calculated according to the theory
outlined in \cite{Hasegawa92}, assuming a sticking coefficient of
unity.

\begin{figure}
  \subfloat[][]{\label{fig:CathplotA}\includegraphics[width=8.5cm]{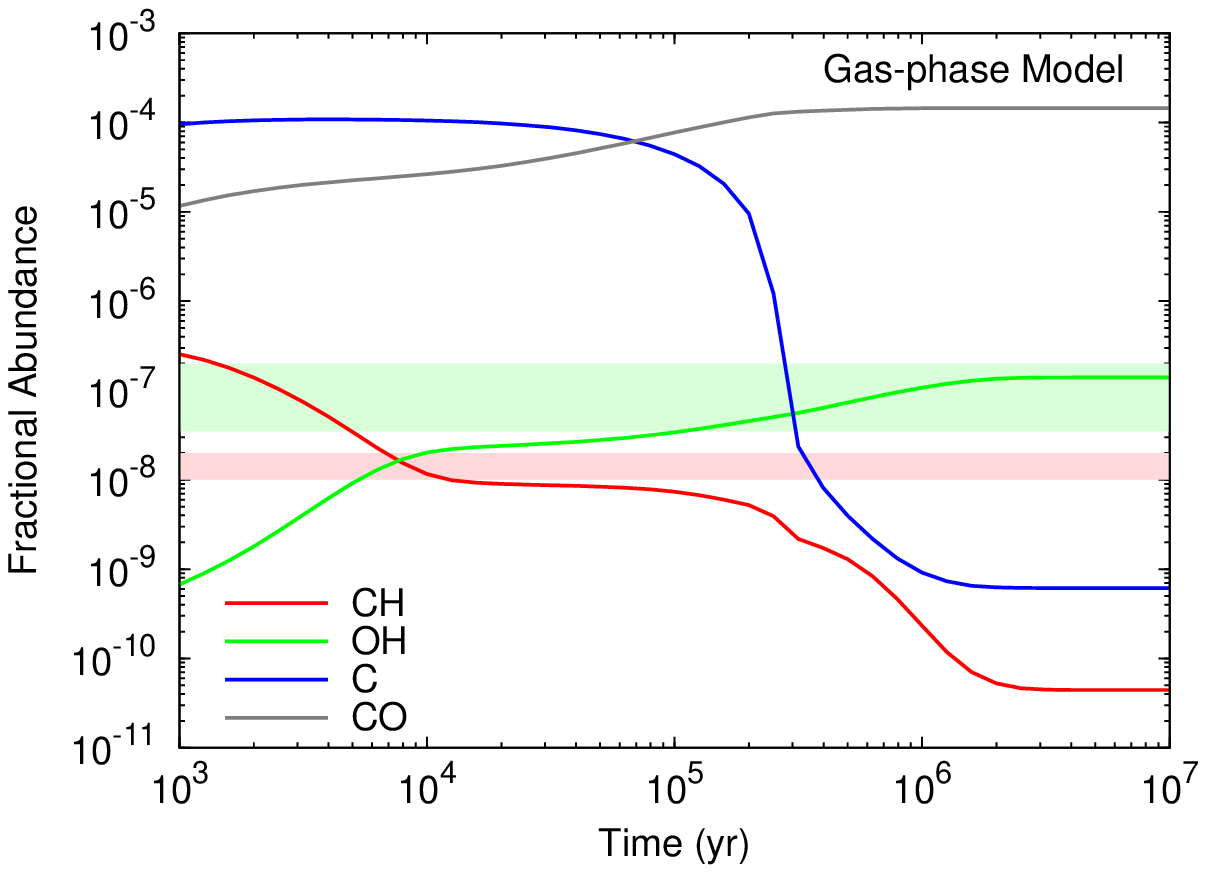}}\\
  \subfloat[][]{\label{fig:CathplotB}\includegraphics[width=8.5cm]{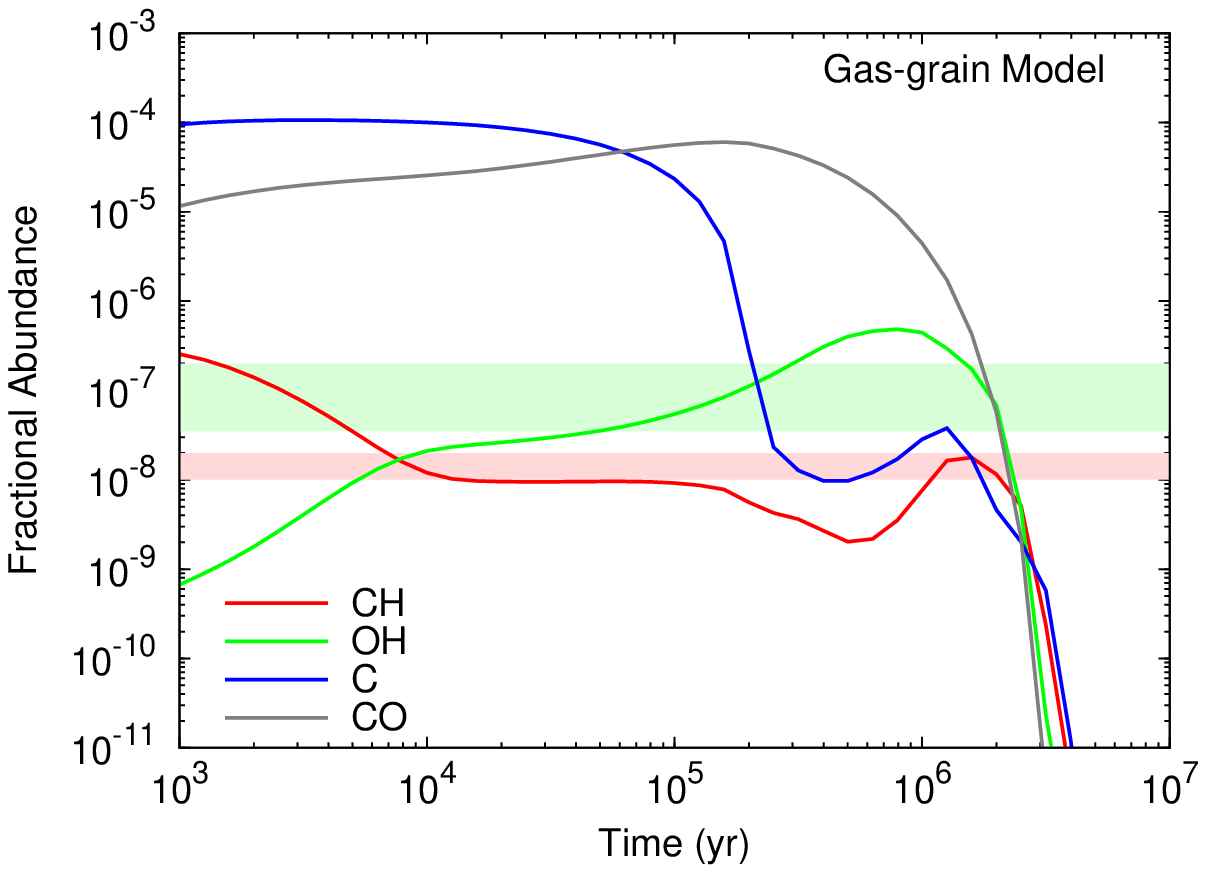}}
  
  \caption{Plots showing the fractional abundances (with respect to H$_2$ number density) 
    of CH and OH as functions of time,
    calculated using a gas-phase chemistry only model (a) and a gas-grain
    model (b) for a \object{TMC-1}-like cloud (see Sect. \ref{sec:chemistry}). 
    Also shown in this Figure are the fractional C and CO abundances. The
    shaded regions represent the observed range of CH and OH fractional
    abundances derived by us along the \object{TMC-1} ridge.}
\end{figure}

The fractional abundances (with respect to H$_2$ number density) 
of CH and OH are plotted in Figs.~\ref{fig:CathplotA}  and \ref{fig:CathplotB}
as functions of time as predicted
by the pure gas-phase (solid lines) and gas-grain (dot-dashed line) models.
We see that the two models agree rather well until an age of
approximately $2\,10^5$ years when the freeze out of molecules starts to take
effect. In both models, the CH and OH fractional abundances 
show an anticorrelation until a time of about $5\,10^5$ years, with CH
continually declining and OH continually increasing. The OH abundance
exceeds that of CH at an early time before $10^4$ years 
and remains thus for the duration of the simulation, 
except until very late stages in the gas-grain model. CO passes
C in abundance in these models around $6-7\,10^4$ years.

Soon after $10^4$ years CH and OH plateau temporarily
(until $\sim 2\,10^5$ year). The relative constancy of the OH/CH abundance
in this time interval can be explained by very similar gas-phase 
production and destruction mechanisms.

According to the present model, the dissociative
recombination of $\rm{H_3O^+}$ is responsible for 81 \% of the OH
production at a cloud age of $10^4$ years and for 89 \% at $10^5$
years. It is only at the relatively
young cloud age of $10^3$ years that dissociative
recombination contributes less than 20 \%.  
At this stage, the model
predicts the reaction
\begin{equation}
{\rm O} + {\rm HCO} \rightarrow {\rm CO} + {\rm OH}
\end{equation}
to be a major process, making up 76 \% of OH formation.  Likewise, the
dissociative recombination of $\rm{CH_3^+}$ is accountable for 72
\% of CH production at $10^4$ years and for 69 \% at $10^3$ years, 
yet contributes only 23 \% at $10^5$ years.  
The main production pathway
at this age is
\begin{equation}
{\rm CH_2} + {\rm H} \rightarrow {\rm CH} + {\rm H_2} \;.
\end{equation}
Since $\rm{CH_2}$ is also likely produced through dissociative 
recombination of $\rm{CH_3^+}$, this ``detour'' 
does not significantly change the overall picture.
All these precursors are formed by 
successive hydrogenation of $\rm{CH^+}$ and $\rm{CH_2^+}$ ions that 
stem from reactions of $\rm{C^+}$ with $\rm{H_2}$ or C with $\rm{H_3^+}$.
Also, $\rm{H_3O^+}$, the main precursor of OH,
has a similar genesis through iterative hydrogenation of $\rm{OH^+}$ formed by protonation 
of O atoms by $\rm{H_3^+}$.

Furthermore, the destruction pathways of CH and OH are very
similar. At a cloud age of $10^5$ years, reactions with C, N and O,
\begin{eqnarray}
{\rm OH} + {\rm C} &\rightarrow& {\rm CO} + {\rm H}\ (31\%)\\
{\rm OH} + {\rm N} &\rightarrow& {\rm NO} + {\rm H}\ (19\%)\\
{\rm OH} + {\rm O} &\rightarrow& {\rm O_2} + {\rm H}\ (49\%)
\end{eqnarray}
are exclusively destroying OH (the
figures in brackets denote the relative importance of each reaction). 
CH is consumed by the following analogous processes:
\begin{eqnarray}
{\rm CH} + {\rm C} &\rightarrow& {\rm C_2} + {\rm H}\ (14\%)\\
{\rm CH} + {\rm N} &\rightarrow& {\rm CN} + {\rm H}\ (21\%)\\
{\rm CH} + {\rm O} &\rightarrow& {\rm CO} + {\rm H}\ (60\%)
\end{eqnarray}
A small proportion is, interestingly, due to the associative ionization 
(reverse dissociative recombination) of CH with O
\begin{equation}
{\rm CH} + {\rm O} \rightarrow {\rm HCO^+} + {\rm e^-}
\end{equation}
which is almost thermoneutral (although if one calculates the Gibbs
energies \citep{Hamberg10}, it becomes endoergic and probably should
be removed from the model). At earlier stages 
(approximately $10^3$ years),
reactions with $\rm{C^+}$,
\begin{eqnarray}
{\rm C^+} + {\rm CH} &\rightarrow& {\rm C_2^+} + {\rm H}\ (42\%)\\
{\rm C^+} + {\rm CH} &\rightarrow& {\rm CH^+} + {\rm C}\ (42\%)
\end{eqnarray}
are more important, because of the higher abundances of $\rm{C^+}$ 
at this time.

The pure gas-phase and gas-grain models start showing large
deviations from each other at times later than $10^5$ years. 
The fractional abundance of CH declines quickly as C is converted into CO, and in the
pure gas-phase model it diminishes 
continually until a steady-state fractional abundance 
of $\approx 3\,10^{-11}$ is reached at a time of about $2\,10^6$ years.
In the gas-grain model around a time of $10^6$ years, the revival of OH and CH 
is due to the depletion of atomic species (C, N and O) as their abundance 
drops rapidly after a time of about $5\,10^5$ years thus limiting their 
effectiveness at destroying the radicals. Beyond $10^6$ years, stable 
species (e.g. CO, N$_2$, O$_2$ and H$_2$O) become locked onto the grain 
surfaces thus depleting the gas-phase of all atoms and molecules.
In the present model,
which includes thermal desorption from grains only, both CH and OH
abundances decrease rapidly after about $2\,10^6$ years. This behaviour is
different from models which include non-thermal desorption
mechanisms (cosmic ray heating, surface photodissociation, exothermic surface
reactions; e.g. \cite{Garrod07}) which predict detectable amounts
of CH and OH at later times.

\subsection{Distributions of {\rm CH} and {\rm OH} in TMC-1}

When looking at our data we see in Fig.~\ref{fig:NCH} a clear
peak in the CH column density in the vicinity of the cyanopolyyne
maximum of \object{TMC-1}, about $3\arcmin$ on its southeastern side (offset
$\Delta x = +14\arcmin$).  
The bump in $N({\rm CH})$ begins at the offset $\Delta
x = +8\arcmin$ which corresponds to the point where our strip enters
the region of greatest intensity on the smoothed SCUBA map
(Fig.~\ref{fig:SCUBAmap228}), and ends at $\Delta x = +18\arcmin$
where we exit the brightest parts.

The diagram in Fig.~\ref{fig:NCH} shows also CH column densities in two velocity
windows, and it is evident that the gas in the velocity range
$5.8-7.5$ km\,s$^{-1}$ is mainly responsible for the peak.  Most
molecular lines observed towards this source, including those of
complex carbon compounds, originate in this velocity range
(e.g. Pratap et al. 1997), and it seems likely that the local CH and
cyanopolyyne peaks are also nearly spatially coincident. Besides the
peak at $\Delta x = +14\arcmin$ the $N({\rm CH})$ in the $V_{\rm LSR}$
range $5.8-7.5$ km\,s$^{-1}$ shows a tendency to increase slightly
from northwest to southeast, all the way until the end of the strip
where the dust column density and $N({\rm CH})$ in the low-velocity
range, $4.5-5.8$ km\,s$^{-1}$ have decreased.

With respect to the fractional CH abundance, $X({\rm CH})$, the CP
peak is not special.  From Fig.~\ref{fig:CHrat} we see that $X({\rm
  CH})$ lies somewhere around $1.0 \, 10^{-8}$ in the northwestern
part of the filament until the offset $\Delta x = +8\arcmin$, and
reaches values of $\sim 2.0\, 10^{-8}$ towards the southeastern end.
This is the only major trend in the variation of $X({\rm CH})$ and
it does not seem to vary in e.g. the vicinity of the high-extinction
cores.

  It is evident from Figures \ref{fig:CHrat}-\ref{fig:CHvOH} that
  the CH abundance is well correlated with that of OH. The fractional
  OH abundance increases from about $2.5\,10^{-7}$ in the northwest to
  about $5.0\,10^{-7}$ in the southeast. Acccording to the 
  chemical
  models discussed in the previous Section, a good correlation between
  OH and CH is expected at youthful cloud ages around $10^4-10^5$
  years because of the very similar gas-phase production and
  destruction mechanisms. During this period, the OH/CH
  abundance ratio in
  the models increases from about 3 to about 10. 
  It may be possible to reproduce the observed OH/CH
  ratio by adjusting the elemental O/C
  ratio and/or the physical parameters of the model.
  We note that during the model ages between $10^4$ and $10^5$ years
the fractional C and CO abundances are within one order of magnitude
from each other, which roughly agrees with the estimates of \cite{Schilke95} 
and \cite{Maezawa99}.

  On the other hand, the {\sl gas-grain
  model} predictions reproduce
  the observed large abundances of CH and OH at later times, around 10$^6$ years. 
  Furthermore, the gas-grain
  model can qualitatively explain the observed spatial variations of
  the CH and OH abundances along the \object{TMC-1} filament if they are
  interpreted in terms of evolution with the chemical age advancing
  from the southeast to the northwest.  After the peak in OH
  in the gas-grain model ($\sim 1.5-2 \,10^6$ years), both CH and OH
  fractional abundances decrease with increasing age, 
  and it is only at this point in the present calculations 
  that this particular trend is seen. 
  Running models including
  a more comprehensive treatment of desorption 
  may shift this period but most likely it would still be
  associated with an advanced stage where accretion onto grain
  surfaces plays a dominant role.

  In view of the rather coarse angular resolution available in the
  present observations, and the fact that both CH and OH are known to
  have large abundances in diffuse gas, it is, however, unlikely that
  observed abundance gradients predominantly reflect conditions in
  depleted cores, which occupy a relatively small fraction of the gas
  encompassed by the telescope beam. A more probable explanation is
  that when moving from the northwestern end of the filament to the
  southeast we see an increasingly thick layer of tenuous gas, which would also agree with
the increasing abundance of neutral carbon towards southeast suggested
by the \ion{C}{I} map of \cite{Maezawa99}. This explanation is furthermore
supported by our chemical models where, 
  if we decrease the gas density, we can see the abundances of both 
  CH and OH increase. In this situation the density dependence
  overcomes the temporal anticorrelation.
  This tendency is probably due to an enhanced production
  by dissociative recombination resulting from the higher fractional
  abundance of electrons in low-density gas.

\cite{Hirahara92} and \cite{Pratap97} estimated by modelling the
excitation of molecular lines that the gas volume density is somewhat
higher near the ammonia peak in the northwestern part of the \object{TMC-1}
filament as compared with the CP peak. The SCUBA maps of
\cite{Nutter08} , reproduced in the present work, show that the CP
peak lies in the area where the dust and gas {\sl column density}
reaches its maximum.  This separation of volume and column density
peaks implies that the southeastern part of the filament possesses a
more extensive envelope of low-density gas than the northwestern part,
and is readily an indication that the stages of dynamical evolution
are different at the two ends of the filament.

The large abundances of CH combined with the presence of the SCUBA
dust intensity maximum in the southeastern parts is consistent with
the presence of the low-density envolope. The fractional abundance
reached at the southeastern end of the strip, $X({\rm CH}) \sim
2.0\, 10^{-8}$, is slightly below half when compared to abundances
observed in diffuse clouds \citep[][e.g. $X(\rm{CH}) \approx 4.3\,
10^{-8}$ in $\zeta$ Per]{Liszt02}.  On the other hand, the relatively
constant abundance $\sim 1.0\, 10^{-8}$ at the offsets $\Delta x <
+8\arcmin$ is likely to correspond to the steady-state value in a
dense core (averaged over a $4\arcmin$ beam). It is, however, somewhat
higher than the abundances predicted by the chemistry model of
\cite{Ruffle97} where the interaction with grains is included. Then
again comparing our calculated abundances to the model presented by
\cite{Flower94}, we see that our abundances are slightly lower than
the values presented in the model results where the depletion of
oxygen $\delta_{\rm O} = 1$ for a cloud with high visual optical
depth. Our results would most likely match Flower's model if a
slightly higher oxygen depletion factor is used as a model where
$\delta_{\rm O}=10$ produces much lower abundances than our
measurements.

\subsection{The structure of the filament in the light of 
the SCUBA map}
In addition to providing us with an estimate of the hydrogen column
density in \object{TMC-1}, the SCUBA maps can be analyzed from the perspective
of examining the physical dimensions of the clumpy structures seen in them.
That is what we intend to do in this subsection.

The clumpfind algorithm \citep{Williams94} was used to identify
clumps, i.e. relatively isolated density enhancements, in the 850
$\rm{\mu m}$ SCUBA map of the filament from \cite{Nutter08}, smoothed
to correspond to a $30\arcsec$ beam.  The filament shows rather small
intensity variations. The best agreement with the visual
inspection is obtained by setting the intensity
threshold to 0.15 Jy/beam (twice the value of the rms noise in the neighbourhood of \object{TMC-1})
and the stepsize to 0.075 Jy/beam ($1 \sigma$). With this setting the algorithm
finds five clumps in the filament; the properties of these clumps are
listed in Table \ref{tab:clumps}. The closest correspondence with the
``Hirahara cores'' from \cite{Hirahara92} is also indicated.
The intensity stepsize used here is half the value recommended 
by \cite{Williams94}. If a $2 \sigma$ step is used, the program 
identifies 3 clumps correponding to the Hirahara cores A+B, C+D, and 
E (extending to the SE tip). The FWHM sizes of the clumps in $x$ and $y$ 
directions and the aspect ratios are calculated in a coordinate system tilted 
by $53.6\degr$ with respect to the R.A. axis  

\begin{table}
\caption{Properties and closest correspondence with Hirahara cores 
of the four clumps found using the clumpfind algorithm on the 850 $\rm{\mu m}$ SCUBA 
intensity map (smoothed to $30\arcsec$ resolution).} 
\label{tab:clumps}      
\centering                           
\begin{tabular}{c c c c c}        
\hline\hline                 
Hirahara	& Mass $[M_\odot]$	& FWHM $X \times Y$ $[\arcmin]$& 
X:Y & $\Sigma$ [$\rm{g/cm^2}$]	\\
\hline                        
A+B      	& 3.9       		& $4.2\times1.5$ & 2.8 	& 0.042 \\
C    		& 4.2           	& $3.6\times1.4$ & 2.6 	& 0.045 \\
D      		& 3.4            	& $2.3\times1.8$ & 1.3 	& 0.045 \\
E	     	& 5.3           	& $3.9\times1.9$ & 2.0 	& 0.041 \\
SE tip          & 1.5                   & $2.9\times1.4$ & 2.2  & 0.032 \\
\hline
\end{tabular}
\end{table}

The total mass of the filament is 18.3 $M_\odot$, and the average
surface density is $<\Sigma> = 0.042$ g\, cm$^{-2}$.  Here we have
assumed a constant dust temperature $T_{\rm d}$=10 K. The previous
determinations of the the gas kinetic temperatures towards various
positions in the cloud, including the CP and NH$_3$ peaks, lie
invariably in the vicinity of 10 K \citep[e.g.][]{Pratap97}, and we
think this value represents well the dust clumps seen at a $30\arcsec$
resolution. The somewhat higher temperatures derived in Sect. \ref{sec:NH2}. by
comparison between the SCUBA 450/850 maps and A$_{\rm V}$, relate to
a $228\arcsec$ beam which encompasses also core envelopes. The clumps
can also be identified as five intensity maxima in
Fig.~\ref{fig:SCUBAmap}.

Even though the locations of the Hirahara cores are always in the
vicinity of a clump on the filament, the actual peaks of the
clumps and the CCS line intensities do not overlap; both of the 850
micron intensity peaks corresponding closest to cores C and D are
shifted about $2\arcmin$ towards each other in relation to the CCS intensity
maxima closest to the two cores. These spatial offsets are larger than
the FWHM of the CCS observations.

Applying the formula quoted by \cite{Hartmann02} for a filament with a
critical line density, one finds that the average surface density
$0.042$ g\, cm$^{-2}$ implies a Jeans length of $\sim 0.16$ pc or
$4\arcmin$ at a distance of 140 pc. The corresponding mass is 2.7
$M_\odot$.  The long-axes of the clumps and the avarage separations
between their centres of mass ($\sim 4.8\arcmin$) are comparabable to
the derived Jeans length. This suggests that the clumps identified
here are formed as a result of thermo-gravitational fragmentation.
The full resolution SCUBA maps and the previous
high-angular resolution molecular line maps of \cite{Langer95}
suggest that the fragmentation has proceeded to still smaller
structures, so that each clump is likely to contain several dense cores.

Finally, we note the overall structure of \object{TMC-1} does not resemble the
model presented in \cite{Hanawa94}, where the fragmentation of
the filament is supposed to have started from one end (from the
northwestern tip in this case). On the SCUBA map we see a roughly
symmetrical structure, the apparent centre lying between cores C and
D. The structure bears resemblance to the final density profiles
resulting from the simulations of \cite{Bastien91}
for the gravitational instability in an elongated isothermal cylinder.

\section{Conclusions}

We report on the detection of a CH abundance gradient in \object{TMC-1} with
X(CH) increasing from $\sim 1.0\, 10^{-8}$ in the northwestern
end of the filament to $\sim 2.0\, 10^{-8}$ in the southeastern end.
The CH column density maximum lies close to cyanopolyyne (CP) peak in
the southeastern part of the filament. As it belongs to the most
simple carbon compounds, CH is likely to reflect the distributions of
\ion{C}{I} or C$^+$.  Using previously published OH
observations, we also show that the fractional abundance of OH increases
strongly from the northwest to the southeast. The OH/CH abundance
ratio lies in the range $\sim 16-20$.

In the northwest, the CH abundance stays remarkably constant at around
$1.0\, 10^{-8}$. The value is clearly higher than the
steady-state values predicted for dense cores by pure gas-phase
chemistry models, but the agreement is better with models including
gas-grain interaction. We performed calculations using the UMIST 
Database for Astrochemistry
reaction network and a homogenous physical model appropriate for sources such as
the \object{TMC-1} ridge.  The best agreement with observations for
the pure gas-phase model was found at cloud ages of several $10^5$ years, but the
predicted CH and OH fractional abundances are a factor of 3-5 lower than the
observed values. Agreement is better with a gas-grain model at this
age. The latter model reproduces the observed abundances at late
stages where molecular depletion is significant. However, in view of
the large beam used in the observations, and the likely
encompassment of low-density gas, it is doubtful that the derived
abundances reflect conditions in depleted cores.

The dust continuum map from SCUBA \cite[][smoothed to the $~4\arcmin$
resolution of the CH observations]{Nutter08} suggests that in the
vicinity of the CP peak the filament has an extensive
envelope, in contrast to the neighbourhood of the NH$_3$
peak which appears to be more compressed.  As the CH and OH
abundances are expected to be anticorrelated with density, their
larger abundances in the southeast are likely to be due to this
low-density envelope. The largest fractional CH and OH abundances are found
at the southeastern end of TMC-1 where we are approaching the "\ion{C}{I}
peak" discovered by \cite{Maezawa99}.

The total H$_2$ column density distribution along the \object{TMC-1} filament
was calculated by combining the SCUBA 850 and 450 $\mu$m maps with an
$A_{\rm V}$ map from 2MASS data. The distribution peaks between the CP
and NH$_3$ maxima. In fact, the high-resolution SCUBA map at 850
$\mu$m shows a rather symmetric structure with the apparent centre
between the Hirahara cores C and D. The sizes and
typical separations of clumps correspond to the Jeans' length in an
isothermal filament with the observed surface density. The main
difference between the two ends of the filament seems to be the
presence of the low-density envelope in the southeast, which probably
is an indication of an early stage of dynamical evolution. As lower
densities result in slower reaction rates, this finding conforms
with the suggestion of several previous studies that the southeastern
end of the cloud is also chemically less evolved than the northeastern
end.

\begin{acknowledgements}
The authors are indebted to several persons who helped in the
preparation of this article. We thank Dr. David Nutter for providing
us with the SCUBA data, and Erik Vigren for his help with the UMIST
model calculations. We are particularly indebted to the anonymous
referee for his/her valuable comments and for drawing our attention to
the possible error in the CH velocities. We are grateful to Prof. Loris
Magnani for providing us with the NRAO 43 m spectra of TMC-1. We thank
Dr. J\"urgen Neidh\"ofer for supplying detailed information on the
Effelsberg telescope receiving system. We are grateful to Prof. Kalevi
Mattila for his support and for very helpful discussions.

A.S., J.H., S.H., and M.J. acknowledge support from the Academy of Finland
through grants 140970, 127015, 201269, and 132291.

Astrophysics at QUB is supported by a grant from the STFC.
\end{acknowledgements}

\bibliographystyle{aa} \bibliography{16079_as}
\end{document}